\documentclass[twocolumn]{aastex63}
\usepackage{comment}
\usepackage{grffile}

\newcommand\name{B14-65666}
\newcommand{\oiii}{[O{\sc iii}]}
\newcommand{\cii}{[C{\sc ii}]}

\newcommand{\lya}{Ly$\alpha$}
\newcommand{\ci}{[C{\sc i}]}

\newcommand{\lsun}{$L_{\rm \odot}$}
\newcommand{\msun}{$M_{\rm \odot}$}
\newcommand{\ltir}{$L_{\rm TIR}$}
\newcommand{\lfir}{$L_{\rm FIR}$}

\newcommand{\lcii}{$L_{\rm [CII]}$}
\newcommand{\loiii}{$L_{\rm [OIII]}$}

\submitjournal{ApJ}
\received{2022 Mar 4}
\accepted{2023 Jun 11}

\shorttitle{Molecular Gas Properties of ``Big Three Dragons''}
\shortauthors{Hashimoto et al.}
\graphicspath{{./}{figures/}}

\begin{document}

\title{Big Three Dragons: Molecular Gas in a Bright Lyman-Break Galaxy at $z=7.15$}

\correspondingauthor{Takuya Hashimoto}
\email{hashimoto.takuya.ga@u.tsukuba.ac.jp}

\author[0000-0002-0898-4038]{Takuya Hashimoto}
\affiliation{Tomonaga Center for the History of the Universe (TCHoU), Faculty of Pure and Applied Sciences, University of Tsukuba, Tsukuba, Ibaraki 305-8571, Japan}

\author[0000-0002-7779-8677]{Akio K. Inoue}
\affiliation{Department of Physics, School of Advanced Science and Engineering, Waseda University, 3-4-1, Okubo, Shinjuku, Tokyo 169-8555}
\affiliation{Waseda Research Institute for Science and Engineering, Faculty of Science and Engineering, Waseda University, 3-4-1, Okubo, Shinjuku, Tokyo 169-8555, Japan}

\author[0000-0001-6958-7856]{Yuma Sugahara}
\affiliation{Waseda Research Institute for Science and Engineering, Faculty of Science and Engineering, Waseda University, 3-4-1, Okubo, Shinjuku, Tokyo 169-8555, Japan}
\affiliation{National Astronomical Observatory of Japan, 2-21-1 Osawa, Mitaka, Tokyo 181-8588, Japan}

\author[0000-0001-7440-8832]{Yoshinobu Fudamoto}
\affiliation{Waseda Research Institute for Science and Engineering, Faculty of Science and Engineering, Waseda University, 3-4-1, Okubo, Shinjuku, Tokyo 169-8555, Japan}
\affiliation{National Astronomical Observatory of Japan, 2-21-1 Osawa, Mitaka, Tokyo 181-8588, Japan}

\author[0000-0001-7201-5066]{Seiji Fujimoto}
\affiliation{Cosmic Dawn Center (DAWN), Copenhagen, Denmark}
\affiliation{Niels Bohr Institute, University of Copenhagen, Jagtvej 128, DK-2200 Copenhagen N, Denmark}

\author[0000-0002-7821-8873]{K.K. Knudsen}
\affiliation{Department of Space, Earth and Environment, Chalmers University of Technology, Onsala Space Observatory, SE-43992 Onsala, Sweden}

\author[0000-0003-3278-2484]{Hiroshi Matsuo}
\affiliation{National Astronomical Observatory of Japan,
2-21-1 Osawa, Mitaka, Tokyo 181-8588, Japan}
\affiliation{Graduate University for Advanced Studies (SOKENDAI), 2-21-1 Osawa, Mitaka, Tokyo 181-8588, Japan}

\author[0000-0003-4807-8117]{Yoichi Tamura}
\affiliation{Department of Physics, Graduate School of Science, Nagoya University, Nagoya 464-8602, Japan}

\author[0000-0002-7738-5290]{Satoshi Yamanaka}
\affiliation{General Education Department, National Institute of Technology, Toba College, 1-1, Ikegami-cho, Toba, Mie 517-8501, Japan}

\author[0000-0002-6047-430X]{Yuichi Harikane}
\affiliation{Institute for Cosmic Ray Research, The University of Tokyo, 5-1-5 Kashiwanoha, Kashiwa, Chiba 277-8582, Japan}
\affiliation{Department of Physics and Astronomy, University College London, Gower Street, London WC1E 6BT, UK}

\author[0000-0002-1234-8229]{Nario Kuno}
\affiliation{Tomonaga Center for the History of the Universe (TCHoU), Faculty of Pure and Applied Sciences, University of Tsukuba, Tsukuba, Ibaraki 305-8571, Japan}

\author[0000-0001-9011-7605]{Yoshiaki Ono}
\affiliation{Institute for Cosmic Ray Research, The University of Tokyo, 5-1-5 Kashiwanoha, Kashiwa, Chiba 277-8582, Japan}

\author[0000-0002-3848-1757]{Dragan Salak}
\affiliation{Institute for the Advancement of Higher Education, Hokkaido University, Kita 17 Nishi 8, Kita-ku, Sapporo, Hokkaido 060-0817, Japan}
\affiliation{Department of Cosmosciences, Graduate School of Science, Hokkaido University, Kita 10 Nishi 8, Kita-ku, Sapporo, Hokkaido 060-0810, Japan}

\author{Nozomi Ishii}
\affiliation{Tomonaga Center for the History of the Universe (TCHoU), Faculty of Pure and Applied Sciences, University of Tsukuba, Tsukuba, Ibaraki 305-8571, Japan}

\begin{abstract}
We report ALMA Band 3 observations of CO(6-5), CO(7-6), and \ci(2-1) in \name\ (``Big Three Dragons''), one of the brightest Lyman-Break Galaxies at $z>7$ in the rest-frame ultraviolet continuum, far-infrared continuum, and emission lines of \oiii\ 88 \micron\ and \cii\ 158 \micron. CO(6-5), CO(7-6), and \ci(2-1), whose $3\sigma$ upper limits on the luminosities are approximately 40 times fainter than the \cii\ luminosity, are all not detected. The \lcii/$L_{\rm CO(6-5)}$ and \lcii/$L_{\rm CO(7-6)}$ ratios are higher than the typical ratios obtained in dusty star-forming galaxies or quasar host galaxies at similar redshifts, and they may suggest a lower gas density in the photodissociated region in \name. By using the (1) \cii\ luminosity, (2) dust mass-to-gas mass ratio, and (3) a dynamical mass estimate, we find that the molecular gas mass ($M_{\rm{mol}}$) is $(0.05-11)\times10^{10}$ \msun. This value is consistent with the upper limit inferred from the nondetection of mid-$J$ CO and \ci(2-1). Despite the large uncertauinty in $M_{\rm mol}$, we estimate a molecular gas-to-stellar mass ratio ($\mu_{\rm{gas}}$) of $0.65-140$ and a gas depletion time ($\tau_{\rm dep}$) of $2.5-550$ Myr; these values are broadly consistent with those of other high-redshift galaxies. \name\ could be an ancestor of a passive galaxy at $z\gtrsim4$ if no gas is fueled from outside the galaxy. 
\end{abstract}



\section{Introduction}\label{sec:intro}

Understanding the properties of molecular gas through cosmic time is an important topic in galaxy formation and evolution, as molecular gas is the fuel for star formation. The molecular gas mass, $M_{\rm mol}$, is often determined from the luminosity of carbon monoxide ($^{12}$C$^{16}$O; hereafter written as simply ``CO’’; e.g. \citealt{bolatto2013}), dust mass (e.g. \citealt{magdis2012}), and radiation from cold dust sensitive to dust mass (e.g. \citealt{scoville2016}). Based on $M_{\rm mol}$ estimates, previous studies have shown that high-redshift ($z\gtrsim2$) star-forming galaxies (SFGs) have (1) higher molecular gas-to-stellar mass ratios ($\mu_{\rm gas} \equiv M_{\rm mol}/M_{\rm *}$) and (2) shorter molecular gas depletion times ($\tau_{\rm dep} \equiv M_{\rm mol}/{\rm SFR}$) than local galaxies (e.g. \citealt{tacconi2020}).

Low-$J$ CO transitions probe the cold and diffuse molecular gas, whereas mid-$J$ transitions\footnote{Hereafter, we refer to CO($J = 6 \rightarrow 5$) and CO($J = 7 \rightarrow 6$) as the mid-$J$ transitions.} probe the warm and dense regions of the molecular gas. 
Based on zoom-in cosmological hydrodynamical simulations implementing radiative transfer calculations, \cite{vallini2019} have shown that galaxies in the epoch of reionization (EoR; $z\gtrsim6 $) have high gas excitation conditions with CO luminosity peaks at an upper rotational level ($J_{\rm u}$) $\approx 6-7$ as a result of their high star-formation surface density and the resulting higher temperature of the giant molecular clouds. These authors show that the sensitivity of the Atacama Large Millimeter/Submillimeter Array (ALMA) telescope is sufficient to detect these mid-$J$ CO transitions in a reasonable amount of integration time.

Observations of the low-$J$ CO transitions in galaxies in the EoR are challenging because these transitions are redshifted to longer radio wavelengths, where instruments are less sensitive. Furthermore, at high redshift, the cosmic microwave background (CMB) has a significant impact upon the CO line emission (e.g. \citealt{sakamoto1999, combes1999, papadopoulos2000, obreschkow2009, da_cunha2013, z.y.zhang2016, tunnard_greve2016}). Firstly, the increased CMB heating leads to a greater population of high rotational levels, thereby boosting higher-$J$ CO luminosities. Secondly, the CMB serves as a stronger background, particularly at the wavelength of the lower-$J$ transitions. As a result, it becomes challenging to observe low-$J$ CO compared to the mid-$J$ transitions at high redshift. Hence, to efficiently detect the molecular gas component, we target the brighter mid-$J$ transitions, which can be observed far more efficiently with sub-/mm facilities.

The \ci\ $^{3}P_{1}$ $\rightarrow$ $^{3}P_{0}$ and \ci\ $^{3}P_{2}$ $\rightarrow$ $^{3}P_{1}$ lines could be more reliable tracers of the bulk of cold gas than mid- and even low-$J$ transitions, particularly under certain conditions (e.g. high cosmic ray flux, low metallicity). For example, \cite{weiss2005, offner2014}, and \cite{glover2015} show that \ci\ is optically thin and traces the surfaces of molecular clouds in a range of environments (e.g. \citealt{papadopoulos2018, shimajiri2013, jiao2019}). 

To date, CO line observations in the EoR were mainly focused on dusty star-forming galaxies (DSFGs) and quasar host galaxies that both have high IR luminosities ($L_{\rm IR} \gtrsim 10^{12} - 10^{13}$ \lsun) and large SFRs $\gtrsim 100-1000$ \msun\ yr$^{-1}$. Among the DSFGs at $z>5$, seven sources were detected in the low-$J$ CO line ($J_{\rm up} =$ 1, 2) 
(\citealt{combes2012, rawle2014, pavesi2018, riechers2013, riechers2020, riechers2021, zavala2022}), and more than 11 sources were detected in the mid-$J$ CO line ($J_{\rm up} \sim6-7$) (\citealt{combes2012, rawle2014, vieira2013, riechers2013, riechers2017, riechers2020, strandet2016, strandet2017, apostolovski2019, zavala2018, casey2019, jin2019, jarugula2021, vieira2022, asboth2016}).
Among the quasar host galaxies at $z\gtrsim6$, at least eight sources were detected in the low-$J$ CO line (\citealt{venemans2017.co.z7, wang2010, wang2011a, wang2016, stefan2015, shao2019}), and more than 25 sources were detected in the mid-$J$ CO line (e.g. \citealt{novak2019, venemans2017.co.z7, venemans2017.co.z6, decarli2022, yang2019, riechers2009, feige.wang2019, li2020.co, walter2003, carilli_walter2013, wang2016, wang2011b}). 
In contrast, among ``normal'' SFGs at $z\sim6$, only one source was detected in the low-$J$ CO line (\citealt{pavesi2019}), and two sources were detected in the CO($J = 6 \rightarrow 5$) line (\citealt{dodorico2018, vieira2022}). 

It is therefore of interest to investigate the nature of the molecular gas in \name\ (``Big Three Dragons’’\footnote{``Big Three Dragons’’ is a hand in a Mahjong game with triplets or quads of all three dragons.}) at $z=7.1520$. This galaxy shows no clear signs of active galactic nucleus (AGN) activity; nonetheless, it is one of the brightest LBGs at $z\gtrsim6$ in the rest-frame ultraviolet (UV) continuum, far-infrared (FIR) continuum, and FIR emission lines of \oiii\ 88 \micron\ and \cii\ 158 \micron\ (\citealt{bowler2014, furusawa2016, hashimoto2019a, sugahara2021}). The large IR and \cii\ luminosities imply the presence of a significant amount of dust and neutral gas, respectively, effectively shielding CO from the UV radiation. Previous studies have also shown that \name\ is an example of the highest-$z$ starburst galaxies owing to a major merger event (\citealt{bowler2017, hashimoto2019a}). Thus, a detailed study of the molecular gas in \name\ may provide information on the connection between mergers, starbursts, the emergence of quasars, and quenching of star formation at high redshift (e.g. \citealt{hopkins2008}).

Herein, we present new ALMA Band 3 observations of \name. Our observational setup efficiently covers CO($J = 6 \rightarrow 5$), CO($J = 7 \rightarrow 6$), and \ci\ $^{3}P_{2}$ $\rightarrow$ $^{3}P_{1}$. \ci\ is highly complementary to mid-$J$ CO; it could trace the bulk of the cold molecular gas component without the need for low-$J$ CO observations.

The rest of this paper is structured as follows. 
In \S \ref{sec:target}, we introduce the target galaxy, \name. In \S \ref{sec:data}, we describe our ALMA Band 3 data. 
In \S \ref{sec:derived_prop}, we calculate the line luminosities and estimate the molecular gas mass in the galaxy.
In \S \ref{sec:results}, we compare \name\ with other high-$z$ objects in terms of the luminosity ratios and interstellar medium (ISM) properties. 
\S \ref{sec:discussion} presents discussions in the context of $\mu_{\rm gas}$ and $\tau_{\rm dep}$. Finally, \S \ref{sec:conclusion} presents our conclusions. 
Throughout this paper, magnitudes are given in the AB system \citep{oke1983}, and we assume a $\Lambda$CDM cosmology with $\Omega_{\small m} = 0.272$, $\Omega_{\small b} = 0.045$, $\Omega_{\small \Lambda} = 0.728$, and $H_{\small 0} = 70.4$ km s$^{-1}$ Mpc$^{-1}$ (\citealt{komatsu2011}). The solar luminosity, \lsun, is $3.839\times10^{33}$ erg s$^{-1}$. Hereafter, we denote CO($J = 6 \rightarrow 5$), CO($J = 7 \rightarrow 6$), and \ci\ $^{3}P_{2}$ $\rightarrow$ $^{3}P_{1}$ as CO(6-5), CO(7-6), and \ci(2-1), respectively.

\section{Our Target: ``Big Three Dragons''}\label{sec:target}

\begin{deluxetable}{lcc}
\tablecaption{Summary of Previous Measurements\label{tab:previous_luminosity}}
\tablewidth{0pt}
\tablehead{
\colhead{Parameters} & \colhead{Measurements} & \colhead{Ref.}
}
\startdata
$L_{\rm UV}$ [\lsun] & $2.0\times 10^{11}$ & B17 \\ 
\loiii\ [\lsun] &  $(3.4\pm0.4)\times 10^{9}$ & H19 \\
\lcii\ [\lsun] &  $(1.1\pm0.1)\times 10^{9}$ & H19 \\
\ltir\ ($T_{\rm d} =$ 40 K, $\beta = 2.0$) [\lsun] &  $4.0 \times 10^{11}$ & S21 \\
\ltir\ ($T_{\rm d} =$ 80 K, $\beta = 1.0$) [\lsun] &  $12.6 \times 10^{11}$ & S21 \\
\lfir\ ($T_{\rm d} =$ 40 K, $\beta = 2.0$) [\lsun] &  $3.1 \times 10^{11}$ & - \\
\lfir\ ($T_{\rm d} =$ 80 K, $\beta = 1.0$) [\lsun] & $5.3 \times 10^{11}$ & -
\enddata
\tablecomments{
The upper limit is $3\sigma$. The total-infrared luminosity, \ltir, and FIR luminosity, \lfir, are estimated by integrating the modified blackbody radiation at $8-1000$ and $42.5-122.5$ \micron, respectively. Following \cite{sugahara2021}, we consider two combinations of ($T_{\rm d}$, $\beta$) = (40 K, 2.0) and (80 K, 1.0). B17, H19, and S21 refer to the studies by \citealt{bowler2017}, \citealt{hashimoto2019a}, and \citealt{sugahara2021}, respectively.
}
\end{deluxetable}

Table \ref{tab:previous_luminosity} summarizes previous observations of the target. The galaxy was discovered by \cite{bowler2014} based on wide-field imaging data of the UltraVISTA survey (e.g. \citealt{mccracken2012}). 
The galaxy has a UV absolute magnitude of $M_{\rm UV} \approx -22.4$, which is $\sim 3-4$ times brighter than the characteristic UV magnitude at $z=7$, $M_{\rm UV}^{*} \approx -21.0$ (e.g. \citealt{bouwens2021}). Subsequently, a high-angular-resolution image taken with the {\it Hubble Space Telescope} ({\it HST}) revealed that B14-65666 comprises two spatially distinct clumps in the rest-frame UV, indicating that the target is experiencing a merger event (\citealt{bowler2017}). 

The spectroscopic redshift of \name\ was obtained with the Faint Object Camera and Spectrograph (FOCAS) on Subaru at $z=7.17$ with \lya\ (\citealt{furusawa2016}). We performed ALMA spectroscopy of \oiii\ 88 \micron\ and \cii\ 158 \micron\ and determined its spectroscopic redshift at $7.1520\pm0.0003$ (\citealt{hashimoto2019a}). Notably, \cite{hashimoto2019a} supported the merger interpretation by showing that \oiii\ and \cii\ can be spatially decomposed into two components associated with the two UV clumps that are kinematically separated by $\approx 150$ km s$^{-1}$. Furthermore, our team (\citealt{hashimoto2019a, sugahara2021}) and \cite{bowler2018, bowler2022} used ALMA to detect the dust continuum emission at $\lambda_{\rm rest} \approx$ 90, 120, and 160 \micron\ with ALMA Band 8, 7, and 6, respectively.

With this large set of multiwavelength line and continuum measurements, \name\ has a well-sampled dust spectral energy distribution (SED). With modified blackbody radiation models for the dust continuum radiation, \cite{sugahara2021} constrained the total-infrared luminosity ($L_{\rm TIR}$; integrated at $8-1000$ \micron) to be 4.0 and 12.6 $\times10^{11}$ \lsun\ with a parameter set of ($T_{\rm d}$, $\beta$) = (40 K, 2.0) and (80 K, 1.0), respectively, where $T_{\rm d}$ and $\beta$ are the dust temperature and emissivity index, respectively. {In the calculation of $L_{\rm TIR}$, the effect of the CMB is corrected following \cite{da_cunha2013}.

\section{ALMA Observations and Data Reduction}\label{sec:data}
We performed ALMA Band 3 observations during Sep $17-22$, 2019, as a Cycle 6 program (ID: 2018.1.01673.S, PI: T. Hashimoto). We used $41-45$ antennas with baseline lengths of $15-2954$ m, resulting in a maximum recoverable scale of $\sim$ 6$\arcsec$. 
Four spectral windows were set at central frequencies of 85.00, 86.88, 97.15, and 98.95 GHz, referred to as SPW1, SPW2, SPW3, and SPW4, respectively. The CO(6-5) line was observed in SPW1, and the CO(7-6) and \ci(2-1) lines were observed in SPW3.
Continuum emission was observed in SPW2 and SPW4. The total on-source exposure time was 3.75 hrs.
The quasar J1008+0029 was used for complex gain calibration. Two quasars, J0854+2006 and J1037-2934, were used for bandpass calibration. The flux was scaled using J0854+2006 and J1037-2934, yielding an absolute accuracy below 5\% in ALMA Band 3.

\begin{deluxetable}{lccc}
\tablecaption{ALMA Band 3 Data\label{tab:data}}
\tablewidth{0pt}
\tablehead{
\colhead{Data} & \colhead{Sensitivity} & \colhead{Beam FWHM} & \colhead{BPA} \\
\colhead{} & \colhead{($\mu$Jy beam$^{-1}$)} & \colhead{($\arcsec$)} & \colhead{($^{\circ}$)}
}
\startdata
{\bf without {\it uv}-taper} \\
\hline
Continuum & 4.6 & 0.46 $\times$ 0.41 & 61 \\
CO(6-5) & 96 & 0.52 $\times$ 0.43 & 66 \\ 
CO(7-6) and \ci(2-1) & 78 & 0.43 $\times$ 0.37 & 66 \\
\hline
{\bf {\it uv}-tapered}\\
\hline
Continuum & 5.3 & 0.82 $\times$ 0.72 & 82 \\
CO(6-5)  & 107 & 0.78 $\times$ 0.68 & 78 \\ 
CO(7-6) and \ci(2-1)  & 93 & 0.78 $\times$ 0.68 & 78
\enddata
\tablecomments{In {\it uv}-tapered data, we adopt taper values of $0\farcs45$, $0\farcs40$, and $0\farcs45$ for the continuum map, CO(6-5) cube, and CO(7-6) cube, respectively. The cube sensitivity is per 50 km s$^{-1}$.}
\end{deluxetable}

The data were reduced and calibrated with the Common Astronomy Software Application (CASA; \citealt{McMullin2007}) pipeline version 5.6.1-8. By using the {\tt tclean} task, we produced maps and cubes with a natural weighting to optimize the point-source sensitivity. 
Table \ref{tab:data} summarizes the resulting resolution and sensitivity of the data. 

Continuum maps were created using all channels that were expected to be line-free. The synthesized beam has a size of $0\farcs46 \times 0\farcs41$ in the full-width at half-maximum (FWHM) and a positional angle (BPA) of 61$^{\circ}$ with a rms value of 4.6 $\mu$Jy beam$^{-1}$. The beam size is smaller than the beam-deconvolved size of the target for the dust continuum and \cii\ emitting region ($\sim 0\farcs8 \times 0\farcs4$ in FWHM, see \citealt{hashimoto2019a}). Therefore, we also created dust continuum maps using a Gaussian taper with a width ranging from $0\farcs0$ to $1\farcs0$. We adopted a taper value of $0\farcs45$ because the resultant beam size ($0\farcs82 \times 0\farcs72$) fully covers the dust continuum emitting region. 

The data probe the dust continuum emission at $\lambda_{\rm rest} \approx 400$ \micron. The left panel of Fig. \ref{fig:maps} shows the nondetection, and by using the {\it uv}-tapered image, we place a $3\sigma$ upper limit of 15.9\,$\mu$Jy on the continuum flux density. The current data is not deep enough to obtain a meaningful constraint on the dust emissivity index.

As the dust continuum was undetected in ALMA Band 3, we created line cubes without performing continuum subtraction. The cubes were rebinned to a velocity resolution of 50\,km\,s$^{-1}$. For SPW1 (SPW3) targeting CO(6-5) [CO(7-6) and \ci(2-1)], we also created a {\it uv}-tapered data cube by using a Gaussian taper with a width of $0\farcs40$ ($0\farcs45$). This cube has a synthesized beam size of $0\farcs78 \times 0\farcs68$ ($0\farcs78 \times 0\farcs68$) and a typical sensitivity of 107 (93) $\mu$Jy beam$^{-1}$.
Hereafter, we use the {\it uv}-tapered maps and cubes unless otherwise specified.

\begin{figure*}[]
\begin{center}
\includegraphics[width=20cm]{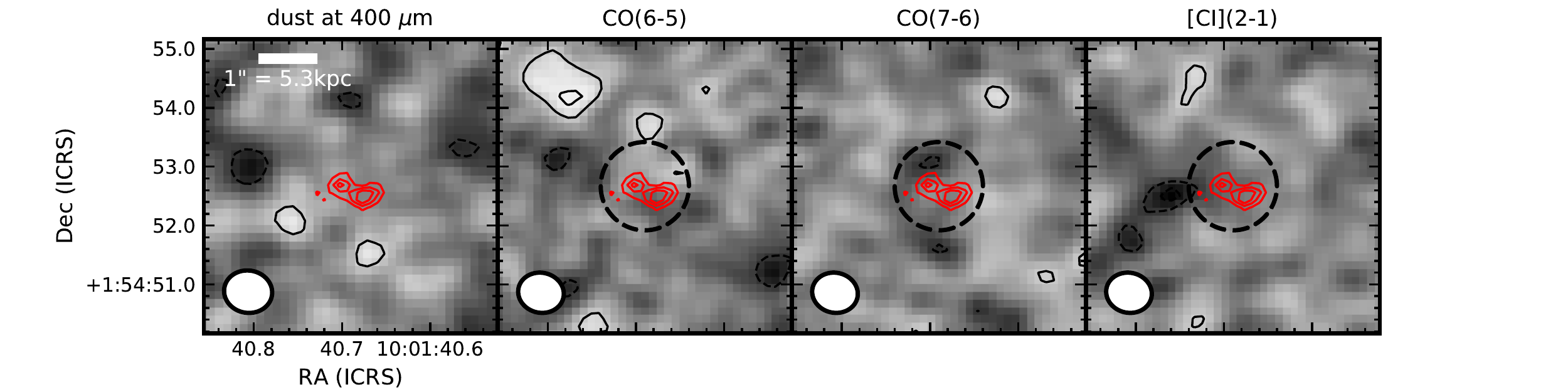}\\
\end{center}
\caption{
From left to right, $5\farcs0 \times 5\farcs0$ cutout images of dust continuum map and integrated intensity maps of CO(6-5), CO(7-6), and \ci(2-1). In each panel, red contours illustrate the morphology in the {\it HST}/WFC3 F140W band that probes the rest-frame UV continuum emission. Black contours are drawn at ($\pm2$, $\pm3$) $\times \sigma$, where the $\sigma$ values are $\approx$ 5.3 $\mu$Jy for the dust continuum map, and 14.9, 14.0, and 13.9 mJy beam$^{-1}$ km s$^{-1}$ for the CO(6-5), CO(7-6), and \ci(2-1) maps, respectively. The black dashed circle at the center shows the $1\farcs5$-diameter aperture used to extract the spectra in Fig. \ref{fig:spectra}. No significant emission has been detected. 
} 
\label{fig:maps}
\end{figure*}

We have searched for the presence of emission lines in the cubes at the position of the target. At $z=7.1520$, the CO(6-5), CO(7-6), and \ci(2-1) emission lines are expected to be at observed frequencies of 84.82, 98.95, and 99.28 GHz, respectively. 
Figure \ref{fig:maps} also shows the integrated intensity maps (i.e., moment 0 maps) of CO(6-5), CO(7-6), and \ci(2-1). In these maps, we integrate the velocity range from $-200$ to $+200$ km s$^{-1}$ with the CASA task {\tt immoments}, which is comparable to the FWHM of \oiii\ and \cii\ (\citealt{hashimoto2019a})\footnote{Several studies show that CO(6-5) and \cii\ have similar FWHMs (e.g. \citealt{wang2013, wang2016, strandet2017, venemans2017.co.z6, zavala2018}). 
}. Figure \ref{fig:spectra} shows the spectra obtained in a $1\farcs5$-diameter aperture centered on the target, where the large aperture size is adopted to capture possible spatially extended CO emission (\citealt{cicone2021}). We conclude that the CO(6-5), CO(7-6), and \ci(2-1) lines are undetected.

\begin{figure}[h]
\begin{center}
\includegraphics[width=7cm]{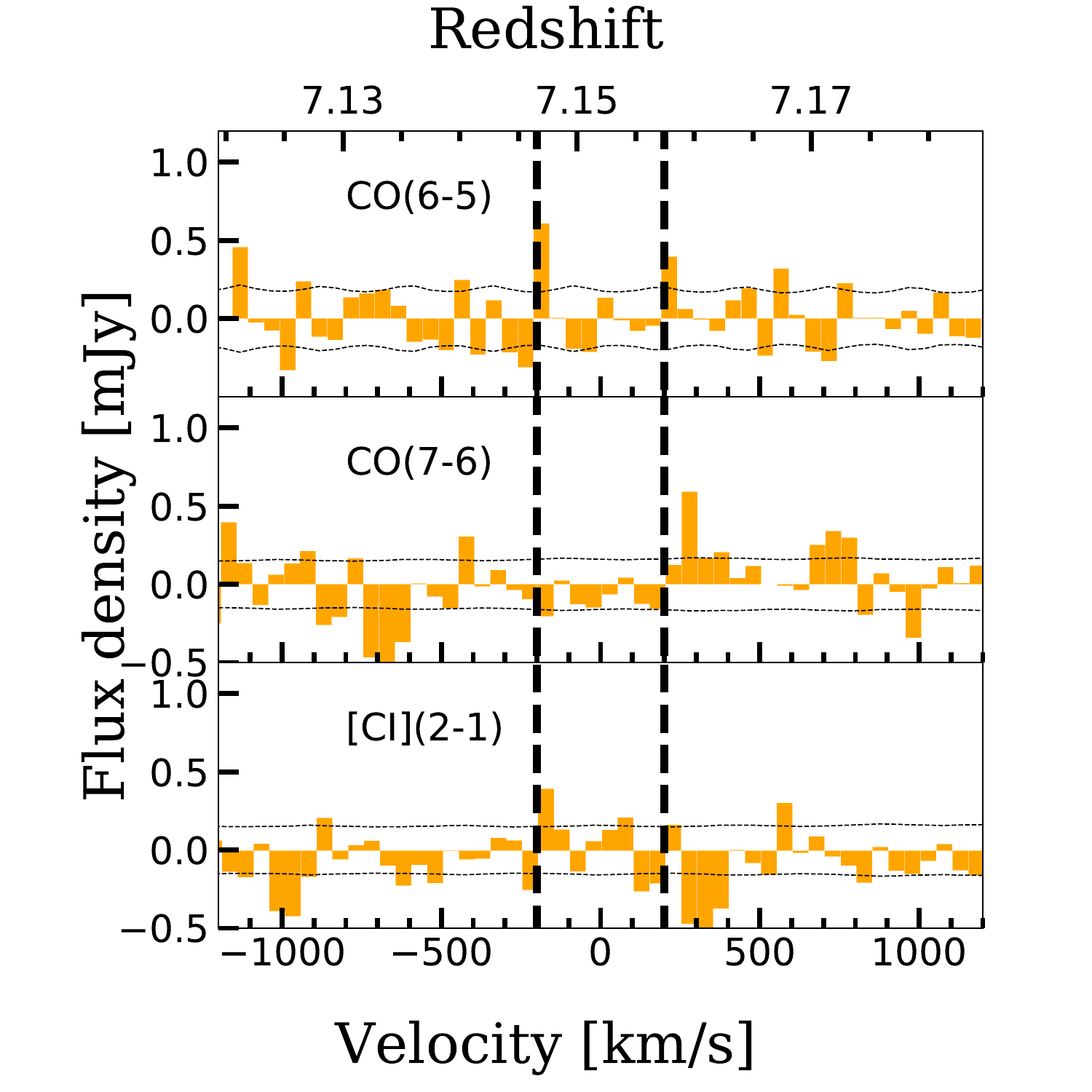}
\end{center}
\caption{Top, middle, and bottom panel shows the spectrum of CO(6-5), CO(7-6), and \ci, respectively (unit: mJy), as extracted from the $1\farcs5$-diameter aperture indicated by the black dashed circle in Fig. \ref{fig:maps}. The black dotted curve shows the noise spectrum. The vertical dashed line shows the velocity range from $-200$ to $+200$ km s$^{-1}$ that is used to create the integrated intensity maps in Fig. \ref{fig:maps}, where the velocity zero-point is defined at $z=7.152$ (\citealt{hashimoto2019a}). The velocity width is set to 50 km s$^{-1}$. \label{fig:spectra}}
\end{figure}

\section{Derived Properties}\label{sec:derived_prop}

\subsection{CO and \ci\ line fluxes}\label{subsec:flux}

From the integrated intensity maps, we obtain the $3\sigma$ upper limits on the velocity-integrated flux, $S_{\rm line} \Delta v$, as 0.0581, 0.0546, and 0.0542 Jy km s$^{-1}$ for CO(6-5), CO(7-6), and \ci(2-1), respectively. Here we assumed that the size of the CO emitting region should not exceed that of the \cii\ emission, which is a better tracer of more extended and multi-phase gas. To obtain the intrinsic line fluxes, we correct for the impact of the CMB.
Following Eq. (32) of \cite{da_cunha2013}, the fraction of the intrinsic line flux observed against the CMB is written as 
\begin{equation}\label{eq:fCMB}
    f_{\rm CMB}= \frac{ S_{\rm \nu/(1+z)}^{J_{\rm u}[{\rm obs \ against \ CMB}]} } { S_{\rm \nu/(1+z)}^{J_{\rm u}[{\rm intrinsic}]} } = 1 - \frac{B_{\rm \nu} [T_{\rm CMB}(z)]}{B_{\rm \nu} [T_{\rm exc}]}, 
\end{equation} 
where $S_{\rm \nu/(1+z)}^{J_{\rm u}[{\rm intrinsic}]}$ and $S_{\rm \nu/(1+z)}^{J_{\rm u}[{\rm obs \ against \ CMB}]}$ represent the intrinsic and observed flux density of the transition of $J_{\rm u}$, respectively. $T_{\rm CMB}(z) = (1+z)\times2.73$ K is the CMB temperature at $z$, and $T_{\rm exc}$ is the excitation temperature in units of K. $B_{\rm \nu}$(T) is the Planck function.
$f_{\rm CMB}$ can be estimated under the assumptions of the local thermal equilibrium (LTE) of molecular clouds and the thermal equilibrium of dust and gas (\citealt{goldsmith2001}). In this case, we can assume $T_{\rm exc} = T_{\rm kin} = T_{\rm dust}$, where $T_{\rm kin}$ is the gas kinetic temperature. We estimate $f_{\rm CMB}$ to be $\sim 0.6-0.9$ using the dust temperature, $T_{\rm dust}\sim40-80$ K, in the target (\citealt{sugahara2021}). 
In the non-LTE case, $f_{\rm CMB}$ depends on a variety of parameters such as $T_{\rm kin}$; number density of H$_{\rm 2}$ molecules, $n_{\rm H_{2}}$; and number density of CO molecules (\citealt{da_cunha2013}). Because the number of CO collisions with H$_{\rm 2}$ becomes small at low $n_{\rm H_{2}}$, the mid- to high-$J$ CO transitions with higher critical densities depart from the LTE case. This leads to $T_{\rm CMB} \sim T_{\rm exc} < T_{\rm kin}$, and it could lead to $f_{\rm CMB}$ as small as 0.1 in the case of $T_{\rm exc}=23$ K (see Fig. 8 in \citealt{combes1999}). In summary, $f_{\rm CMB}$ is highly uncertain, ranging from $\sim 0.1-0.9$ at $T_{\rm exc} = 23-80$ K. 

\begin{deluxetable}{lccc}
\tablecaption{Summary of Measurements\label{tab:measurements}}
\tablewidth{0pt}
\tablehead{
\colhead{Parameters} & \colhead{CO(6-5)} & \colhead{CO(7-6)} & \colhead{\ci(2-1)} 
}
\decimalcolnumbers
\startdata
$S_{\rm line} \Delta v$  & $<$ 0.0581 &  $<$ 0.0546 & $<$ 0.0542  \\
$S_{\rm line} \Delta v$(corr.) &  $< \frac{0.0581 }{f_{\rm CMB}}$ & $ < \frac{0.0546}{f_{\rm CMB}}$ & $ < \frac{0.0542}{f_{\rm CMB}}$ \\
\hline
$L_{\rm line}$ & $<$ 2.70 &  $<$ 2.96  & $<$ 2.95 \\ 
$L_{\rm line}$(corr.)  &  $< \frac{2.70}{f_{\rm CMB}}$ & $< \frac{2.96}{f_{\rm CMB}}$  &  $< \frac{2.95}{f_{\rm CMB}}$  \\ 
\hline
$L'_{\rm line}$ & $<$ 2.55  & $<$ 1.77  & $<$ 1.74  \\ 
$L'_{\rm line}$(corr.) & $< \frac{2.55}{f_{\rm CMB}}$  & $< \frac{1.77}{f_{\rm CMB}}$ & $< \frac{1.74}{f_{\rm CMB}}$  \\ 
\hline 
\loiii/$L_{\rm line}$ & $>$ 125  & $>$ 115  & $>$ 115 \\
\loiii/$L_{\rm line}$(corr.) &  $> 125 f_{\rm CMB}$  & $> 115 f_{\rm CMB}$  &  $> 115 f_{\rm CMB}$ \\
\hline 
\lcii/$L_{\rm line}$ &  $>$ 41  & $>$ 37  &  $>$ 37  \\
\lcii/$L_{\rm line}$(corr.) &  $> 41 f_{\rm CMB}$  &  $>37 f_{\rm CMB}$  & $> 37 f_{\rm CMB}$
\enddata
\tablecomments{
The limits correspond to $3\sigma$. 
$S_{\rm line} \Delta_{v}$ is the line flux in units of Jy km s$^{-1}$. $L_{\rm line}$ and $L'_{\rm line}$ are the line luminosities in units of $10^{7}$ \lsun\ and $10^{9}$ K km s$^{-1}$ pc$^{2}$, respectively. 
The CMB-corrected values are shown with ``(corr.)'', where $f_{\rm CMB}$ ranges from 0.1 to 0.9 (see the main text). 
}
\end{deluxetable}

\subsection{Upper limits on CO and \ci\ line luminosities}\label{subsec:line_luminosity}
We obtain the $3\sigma$ upper limits on two types of line luminosities (\citealt{solomon1992, carilli_walter2013}), which are summarized in Table \ref{tab:measurements}.
The first one, $L_{\rm line}$ in units of \lsun, is written as 
\begin{equation}\label{eq:Lline}
L_{\rm line} = 1.04 \times 10^{-3} \times S_{\rm line} \Delta v D_{\rm L}^{2} \nu_{\rm obs},
\end{equation}
where $S_{\rm line} \Delta v$ is the velocity-integrated flux in units of Jy km s$^{-1}$, $D_{\rm L}$ is the luminosity distance in Mpc, and $\nu_{\rm obs}$ is the observed frequency in GHz. 
The second one, $L_{\rm line}^{'}$, corresponds to the area-integrated brightness in units of K km s$^{-1}$ pc$^{2}$, and it is written as 
\begin{equation}\label{eq:Lpline}
L_{\rm line}^{'} = 3.25 \times 10^{7} \times S_{\rm line} \Delta v \frac{D_{\rm L}^{2}}{(1+z)^3 \nu_{\rm obs}^2}.
\end{equation}
With $S_{\rm line} \Delta v$ (Table \ref{tab:measurements}), the $3\sigma$ upper limits on $L_{\rm line}$ ($L_{\rm line}^{'}$) are $\frac{2.70}{f_{\rm CMB}}$, $\frac{2.96}{f_{\rm CMB}}$, and $\frac{2.95}{f_{\rm CMB}}$ $\times 10^{7}$ \lsun\ ($\frac{2.55}{f_{\rm CMB}}$, $\frac{1.77}{f_{\rm CMB}}$, and $\frac{1.74}{f_{\rm CMB}}$ $\times 10^{9}$ K km s$^{-1}$ pc$^{2}$) for CO(6-5), CO(7-6), and \ci(2-1), respectively.

\subsection{Molecular gas mass estimates}\label{subsec:mol_mass}
We estimate the molecular gas mass of \name. In light of the rich dataset, we adopt five techniques as summarized in Table \ref{tab:gas_mass}. 

\subsubsection{Estimates with CO(6-5) and CO(7-6)}\label{subsec:mol_mass_CO}
The molecular gas mass is estimated with CO lines as 
\begin{equation}
\frac{ M_{\rm mol}^{\rm CO} }{ M_{\odot} } = \alpha_{\rm CO} \ r_{J1}^{-1} \ L^{'}_{{\rm CO}_{J \rightarrow J-1}}, \label{eq:co_mol} 
\end{equation}
where $\alpha_{\rm CO}$ is the CO-to-H$_{2}$ conversion factor in units of $M_{\rm \odot}$ (K km s$^{-1}$)$^{-1}$, and $r_{J1}$ is the excitation correction factor defined as 
\begin{equation}
r_{J1} = \frac{ {L^{'}_{\rm CO}}_{J \rightarrow J-1} } { {L^{'}_{\rm CO}}_{1 \rightarrow 0} } 
= \frac{ {I_{\rm CO}}_{J \rightarrow J-1}} { {I_{\rm CO}}_{1 \rightarrow 0} } \ \frac{1}{J^{2}} \label{eq:excitation}. 
\end{equation}

We use Eq. (19) of \cite{narayanan2014}, who have shown that the CO excitation ladders can be parametrized with $\Sigma_{\rm SFR}$ based on simulations of disc galaxies combined with CO line radiative transfer calculations. With $\Sigma_{\rm SFR} = 20.6^{+11.4}_{-7.6}$ \msun\ yr$^{-1}$ kpc$^{-2}$ obtained for the target\footnote{The target has SFR $= 200^{+82}_{-32}$ \msun\ yr$^{-1}$ and the \oiii\ beam-deconvolved size of $(3.8\pm0.5) \times (2.2\pm0.6)$ kpc$^{2}$ in FWHM (\citealt{hashimoto2019a}). The $\Sigma_{\rm SFR}$ value is calculated as $\frac{\rm SFR}{2\pi r^{2}}$, where $r$ is the half-light radius.}, $r_{61} = 0.28^{+0.04}_{-0.02}$ and $r_{71} = 0.17^{+0.02}_{-0.02}$. 
With the upper limit on the CO(7-6) luminosity (Table \ref{tab:measurements}), we obtain 
$L^{'}_{{\rm CO}_{J = 1 \rightarrow 0}} < \frac{1.2 \times10^{10}  }{f_{\rm CMB}}$ K km s$^{-1}$ pc$^{2}$ ($3\sigma$). 

Previous observational studies (e.g. \citealt{leroy2011, shi2016}) as well as theoretical ones (e.g. \citealt{wolfire2010, narayanan2012}) show that $\alpha_{\rm CO}$ increases at lower gas-phase metallicity as a result of increased CO photodissociation. In this study, we adopt the conversion factor of \cite{tacconi2018} (their Eq. (2)) that is a function of the gas-phase metallicity. 
The gas-phase metallicity of \name\ is estimated to be $0.4^{+0.4}_{-0.2}\ Z_{\rm \odot}$ based on SED fits by taking into account the multiwavelength data ranging from rest-frame UV to FIR (\citealt{hashimoto2019a}). 
With a broad range of $0.2-0.8 Z_{\rm \odot}$ (i.e., 12+log(O/H) = $8.0-8.6$), $\alpha_{\rm CO} \approx 5-25$ \msun\ (K km s$^{-1}$ pc$^{2}$)$^{-1}$. 

With  $\alpha_{\rm CO} = 25$ \msun\ (K km s$^{-1}$ pc$^{2}$)$^{-1}$ and the $3\sigma$ upper limit on $L^{'}_{{\rm CO}_{J = 1 \rightarrow 0}}$, we estimate the molecular gas mass to be $M_{\rm mol}^{\rm CO} < \frac{3.0 \times 10^{11}}{f_{\rm CMB}}$ \msun\ ($3\sigma$). With $f_{\rm CMB} \sim 0.1-0.9$, the $3\sigma$ upper limit becomes $\sim  (4 - 30) \times 10^{11}$ \msun.  Similarly, we obtain the $3\sigma$ upper limit of $(3 - 25)  \times 10^{11}$ \msun\ from CO(6-5).

\subsubsection{Estimate with \ci(2-1)}\label{subsubsec:mol_mass_CI}
The neutral carbon mass, $M_{\rm CI}$, can be obtained from the \ci\ luminosity and $T_{\rm exc}$. We estimate $M_{\rm CI}$ following \cite{weiss2003} as
\begin{eqnarray}
\frac{M_{\rm CI}}{M_{\rm \odot}} = 4.566\times10^{-4} Q(T_{\rm exc}) \frac{1}{5} e^{62.5/T_{\rm exc}} \frac{L^{'}_{\rm [CI](2-1)}}{f_{\rm CMB}}, 
\end{eqnarray}
where $Q(T_{\rm exc}) = 1 + 3e^{-T_{\rm 1}/T_{\rm exc}} + 5e^{-T_{\rm 2}/T_{\rm exc}}$ is the \ci\ partition function, and $T_{\rm 1} = $ 23.6 K and $T_{\rm 2} = $ 62.5 K is the temperature of each transition from the ground state. 
By using $f_{\rm CMB} \sim 0.1-0.9$ at $T_{\rm exc} \sim 23 - 80$ K and the CMB-corrected luminosity of \ci(2-1), we obtain $M_{\rm CI} < (2.2 - 59) \times 10^{6}$ \msun\ ($3\sigma$). 
Assuming the abundance ratio of $X$[{\sc Ci}]/$X$[H$_{2}$] $\sim 1.6 \times 10^{-5}$ as obtained in $z\sim1$ main-sequence galaxies (\citealt{valentino2018}), the \ci(2-1)-based molecular gas mass is $M_{\rm mol}^{\rm [CI]} < (3.0 - 82) \times 10^{10}$ \msun\ ($3\sigma$), where the contribution of helium is included. 
\cite{heintz&watson2020} have revealed that the mass conversion factor of the \ci(1-0) transition, $\alpha_{\rm [CI](1-0)} \equiv M_{\rm mol}/L^{'}_{\rm [CI](1-0)}$, depends on the metallicity based on observations of \ci($J=1$)\footnote{$J$ refers to the total angular momentum quantum number for this transition.} absorption lines in the rest-frame UV toward a sample of gamma-ray burst (GRB) and quasar absorption systems at $z\sim1.9-3.4$. $\alpha_{\rm [CI](1-0)}$ becomes approximately 10 times higher at $0.2 Z_{\rm \odot}$ than at $Z_{\rm \odot}$. If we assume that the mass conversion factor of \ci(2-1) similarly changes with metallicity, our upper limits can be higher by a factor of 10, $M_{\rm mol}^{\rm [CI]} \lesssim (3.0 - 82) \times 10^{11}$ \msun ($3\sigma$) (Table \ref{tab:gas_mass}). 

\subsubsection{Estimate with \cii\ 158 \micron}\label{subsubsec:mol_mass_CII}

The \cii\ 158 \micron\ line can also be used to probe the molecular gas mass (\citealt{zanella2018, madden2020, dessauges-zavadsky2020} and references therein).
We use the conversion factor $\alpha_{\rm [CII]}$ of \cite{madden2020} (see their Eq. (5)) that is appropriate for metal-poor galaxies.

We apply two corrections to the \cii\ luminosity. First, we remove the \cii\ contribution originating from the {\sc Hii} region, 
although it becomes negligible in galaxies with, for example, low-$Z$ (e.g. \citealt{croxall2017}). 
From the metallicity of the target and Fig. 9 of \cite{cormier2019}, we estimate the contribution from the {\sc Hii} region to be $\approx30$\%.
Second, we correct for the CMB impact to \cii. Based on semi-analytical model of galaxy formation combined with photoionization modelling, \cite{lagache2018} have shown that the \cii\ luminosity can be reduced by 0.3 dex ($f_{\rm CMB} = 0.5$) at $z=7$ (see their Figure 4) in the case of a photodissociated region (PDR) with the hydrogen nuclei density of log($n$(H)) $= 2.4$ irradiated by the incident FUV radiation field of $3.2\times10^{3} G_{\rm 0}$, where $G_{\rm 0}$ is the Habing Field in unit, $1.6 \times 10^{-3}$ erg cm$^{-2}$ s$^{-1}$ (\citealt{habing1968}). These PDR parameters are similar to those obtained in $z\sim3-4$ DSFGs (e.g. \citealt{wardlow2017}), and are not improbable for \name. Similarly, based on the cosmological hydrodynamic simulations combined with radiative transfer calculations, \cite{vallini2015} have also modeled the \cii\ emission at $z\sim7$ taking the CMB effect into account. These authors find that the \cii\ emission from the PDR is not severely impacted by the CMB effect, only up to 20\%\ (see similar results in \citealt{kohandel2019}). Given the uncertainty, we assume $f_{\rm CMB}=0.5-1.0$ in \name.  The intrinsic \cii\ luminosity from the molecular gas is $\approx (7.7 - 17) \times 10^{8}$ \lsun. We thus obtain $M_{\rm mol}^{\rm [CII]} \approx (5.4-23) \times 10^{10}$ \msun, where we include the helium contribution and take into account a standard deviation of 0.14 dex in the relation.

\subsubsection{Estimate with dust continuum}\label{subsubsec:dust}

We estimate the gas mass based on $M_{\rm d}$ and the metallicity-dependent dust-to-gas ratio (DGR; \citealt{remy-ruyer2014, li2019}).
With the prescription of \cite{li2019} derived from cosmological hydrodynamical simulations implementing the process of dust production, growth, and destruction (see their Eq. (9)), we obtain DGR $\approx (1.8-53) \times 10^{-4}$ at the metallicity of the target. 
Combined with the dust mass of the target, log($M_{\rm dust}/M_{\rm *}$) $\approx 6.4-7.5$ (\citealt{sugahara2021}), we estimate the (molecular + atomic) gas mass to be $M_{\rm gas}^{\rm dust} \approx (0.05-17) \times 10^{10}$ \msun. If we assume that gas is predominantly in the molecular phase (\citealt{riechers2013}), this can be regarded as the molecular gas mass. 

\subsubsection{Upper limit with dynamical mass}\label{subsubsec:mol_mass_dynamical}

We calculate an upper limit on $M_{\rm mol}$ from the dynamical mass, $M_{\rm dyn}$, subtracted by the stellar mass contribution. \cite{hashimoto2019a} obtained $M_{\rm dyn}$ of two individual clumps of \name\ based on the line width and beam-deconvolved size of \cii\ 158 \micron\ under the assumption of the virial theorem. The dynamical mass of the whole system is estimated to be $M_{\rm dyn} = (8.8\pm1.9) \times 10^{10}$ \msun, where the error only considers the measurement uncertainties. 
With a stellar mass obtained from SED fitting ($M_{\rm *} = 7.7^{+1.0}_{-0.8} \times 10^{8}$ \msun; \citealt{hashimoto2019a}), we obtain a conservative upper limit on $M_{\rm mol}$ to be $\sim 11 \times 10^{10}$ \msun.

To summarize, by combining the $M_{\rm mol}$ estimates from the \cii\ luminosity, dust mass, and dynamical mass, we obtain $M_{\rm mol} = (0.05 - 11) \times 10^{10}$ \msun, which is consistent with the upper limits on $M_{\rm mol}$ inferred from the nondetections of mid-$J$ CO and \ci(2-1). Although the Band 3 observations were conducted to constrain $M_{\rm mol}$ in \name, we note that the tightest constraint on $M_{\rm mol}$ comes from the previous observations of dust and \cii\ 158 \micron, not from mid-$J$ CO or \ci(2-1), due to the insufficient sensitivity of the Band 3 observations. Future deeper Band 3 observations are crucial to better constrain $M_{\rm mol}$ with mid-$J$ CO or \ci(2-1).

\begin{deluxetable}{lc}[t]
\tablecaption{Molecular Gas Mass \label{tab:gas_mass}}
\tablewidth{0pt}
\tablehead{
\colhead{Method} & \colhead{$M_{\rm mol}$}  \\ 
\colhead{} & \colhead{$10^{10}$ \msun}
}
\startdata
CO(6-5) & $<(40-300)$ \\
CO(7-6) & $<(30-250)$ \\
\ci(2-1) & $<(30-820)$ \\ 
\cii\ 158 \micron & $5.4-23$ \\ 
dust & $0.05-17$  \\
dynamical mass & $< 11$ 
\enddata
\tablecomments{
The estimates based on mid-$J$ CO and \ci(2-1) are $3\sigma$ upper limits, where the values in the parenthesis reflect the uncertainty in the CMB correction. The estimate based on the dynamical mass ($M_{\rm dyn}$) provides the upper limit. 
}
\end{deluxetable}

\section{Results}\label{sec:results}
\subsection{Luminosity comparisons}\label{subsec:luminosity_ratios}

\begin{figure*}[]
\begin{center}
\hspace*{-3cm}
\includegraphics[width=22cm]{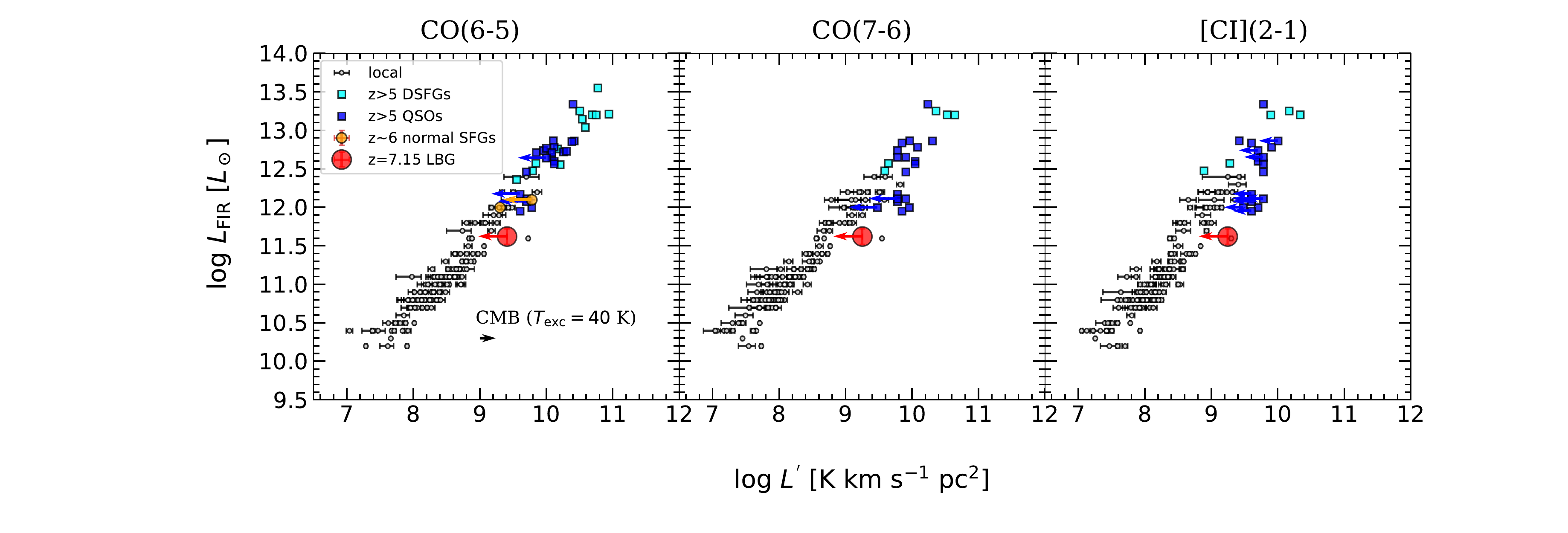}
\end{center}
\vspace*{-1cm}
\caption{
Far-infrared luminosity defined in the range of $42.5-122.5$ \micron\ plotted against the line luminosity. The red circle, cyan squares, blue squares, and orange circles show the data points of \name, $z>5$ DSFGs, quasar host galaxies, and normal SFGs (see the text for the details of the literature sample), respectively, where the upper limits correspond to $3\sigma$. For the detections at $z>5$, the typical significance levels are 7, 6, and 4 for CO(6-5), CO(7-6), and \ci(2-1), corresponding to the line luminosity uncertainties of 0.1, 0.1, and 0.2 dex, respectively. The line luminosities are before the CMB correction. The black arrow in the left panel shows the impact of CMB at $T_{\rm exc}=40$ K; it shifts the data points toward higher line luminosities by $\sim0.2$ dex at $z=7$.
Small open circles show a compilation of local objects, including SFGs, AGNs, and U/LIRGs observed with {\it Herschel}/SPIRE (\citealt{kamenetzky2016}), where objects with $>3 \sigma$ detections are plotted. 
\label{fig:pLline_LFIR}}
\end{figure*} 

\subsubsection{mid-$J$ CO and \ci\ vs. far-infrared luminosity}\label{subsubsec:luminosity_ratios1}

In the local universe, a compiled sample of SFGs, AGNs, and U/LIRGs observed by {\it Herschel}/SPIRE shows a positive correlation between the mid-$J$ CO and the \ci(2-1) line luminosities and FIR luminosity, \lfir\ (e.g. \citealt{kamenetzky2016}). 
Figure \ref{fig:pLline_LFIR} shows a comparison of \name\ with the local objects (\citealt{kamenetzky2016}). The FIR luminosity of \name\ is calculated by integrating the modified blackbody radiation at $42.5-122.5$ \micron, where the CMB effect is corrected following \cite{da_cunha2013} (Table \ref{tab:previous_luminosity}). $z\sim5-7$ DSFGs (\citealt{jarugula2021, riechers2013, zavala2018, casey2019, vieira2022, apostolovski2019, riechers2017, riechers2020, combes2012}), quasar host galaxies (\citealt{novak2019, venemans2017.co.z7, venemans2017.co.z6, decarli2022, yang2019, riechers2009, feige.wang2019}), as well as $z\sim6$ normal SFGs (\citealt{dodorico2018, vieira2022}) are also plotted, where the lensing magnification is corrected when necessary. Note that the number of data points differs in each transition. Although we show the line luminosities not corrected for the impact of the CMB, it shifts the line luminosities toward higher values by 0.2 dex at $T_{\rm exc} = 40$ K at $z=7$, as indicated by a black arrow in the left panel. 
Figure \ref{fig:pLline_LFIR} shows that high-$z$ sources also seem to follow the correlations. This might imply that the CMB effect may not be severe even at high redshift, although this could be due to a bias towards bright DSFGs and quasar host galaxies with higher $T_{\rm exc}$.

The data points of \name\ for the first time place constraints on the line luminosities at log(\lfir/\lsun) $<12.0$ at $z>6$. Nevertheless, the upper limits are loose at a given \lfir, especially when the CMB impact is taken into account. This indicates that the nondetection of the lines can be explained by the insufficient sensitivity of the observations.

\subsubsection{mid-$J$ CO and \ci\ vs. \cii\ 158 \micron\ luminosity}\label{subsubsec:luminosity_ratios2}
 
\begin{figure*}[]
\begin{center}
\hspace*{-3cm}
\includegraphics[width=22cm]{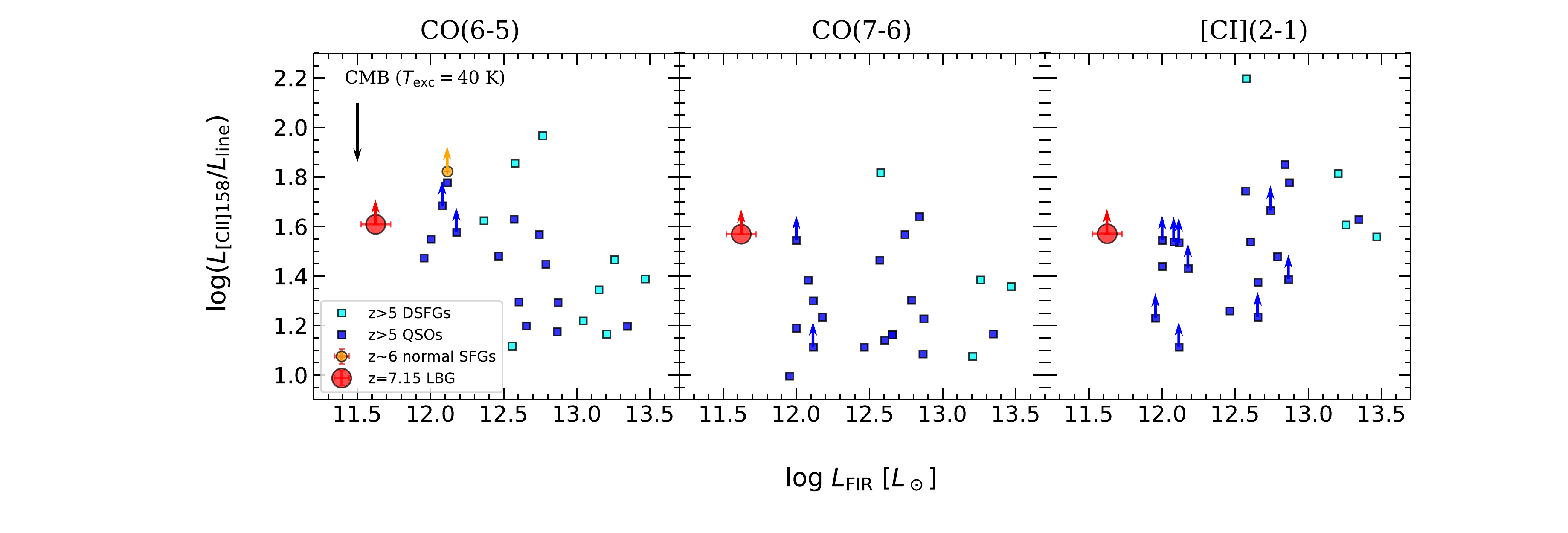}
\end{center}
\vspace*{-1cm}
\caption{
From left to right, \lcii/$L_{\rm CO(6-5)}$, \lcii/$L_{\rm CO(7-6)}$, and \lcii/$L_{\rm [CI](2-1)}$ are plotted against \lfir. The red circle, cyan squares, blue squares, and orange circles show the data points of \name, $z>5$ DSFGs, quasar host galaxies, and normal SFGs, respectively, where the upper limits correspond to $3\sigma$. The line luminosities are before the CMB correction. The black arrow in the left panel shows the impact of the CMB on the lines at $T_{\rm exc}=40$ K; it shifts the data points toward lower line luminosity ratios by $\sim -0.2$ dex at $z=7$.
\label{fig:LFIR_LCII_Lline}}
\end{figure*}

Figure \ref{fig:LFIR_LCII_Lline} shows plots of \lcii/$L_{\rm CO(6-5)}$, \lcii/$L_{\rm CO(7-6)}$, and \lcii/$L_{\rm [CI](2-1)}$ against \lfir. It also shows plots of DSFGs, quasar host galaxies, and normal SFGs at $z>5$, with the luminosity measurements as in Fig. \ref{fig:pLline_LFIR}. The luminosities are before the CMB correction.

\name\ has line luminosity ratios $\gtrsim 40$ ($3\sigma$). If we focus on \lcii/$L_{\rm CO(7-6)}$, the lower limit is three times higher than the predicted value of $\sim13$ for a simulated galaxy at $z=6$ in \cite{vallini2019}, namely, ``Alth\ae a,'' for which $M_{\rm *} \approx 10^{10}\ M_{\rm \odot}$, SFR $\approx 100$ \msun\ yr$^{-1}$, and $Z\sim 0.5\ Z_{\rm \odot}$. \name\ has similar SFR and metallicity values; however, its stellar mass is approximately one order of magnitude lower than that of Alth\ae a. Although only five (three) objects have \lcii/$L_{\rm CO(6-5)}$ (\lcii/$L_{\rm CO(7-6)}$) measurements higher than \name, its interpretation is complicated owing to the large uncertainty in $f_{\rm CMB}$. If \name\ has low $n_{\rm H_{2}}$ and/or gas temperature compared with those of DSFGs or quasar host galaxies at similar redshifts, $f_{\rm CMB}$ in \name\ becomes small, making the lower limits of \name\ more consistent with the typical values in high-$z$ DSFGs and quasar host galaxies. Because a large fraction of the data points in Figure \ref{fig:LFIR_LCII_Lline} comes from quasar host galaxies, the difficulty in measuring their stellar mass, size, and SFR surface densities also prevents us to further examine the physical origins of the higher luminosity ratios in \name\ than other EoR sources. The situation will be improved by {\it the James Webb Space Telescope} that provides these measurements in quasar host galaxies.

In summary, the current data is insufficient to examine the difference in the CMB-corrected $L_{\rm [CII]}/L_{\rm line}$ in \name\ and other high-$z$ objects. The results also imply that care must be taken when comparing the luminosity ratios of galaxies in the EoR.

\subsection{PDR modelling}\label{subsec:pdr_model}

The luminosity ratios are useful to examine properties of the interstellar medium (ISM: e.g. \citealt{kaufman2006, pound_wolfire2008}), although the impact of the CMB makes the interpretation complicated, as stated in \S \ref{subsubsec:luminosity_ratios2}. 
The \lcii/$L_{\rm [CI](2-1)}$ luminosity ratio is sensitive to the heating source of the ISM (\citealt{meijerink2007}). The high ratio, $\gtrsim 40$, excludes the possibility that the lines are heated by the X-ray dominated regions, where \lcii/$L_{\rm [CI](2-1)} \lesssim6$ is expected. We thus compare the line ratios of \name\ to the model predictions of PDR Toolbox (version wk2020) to place constraints on the physical properties of the PDRs. The model assumes a geometry of infinite plane slabs of hydrogen characterized by the hydrogen nuclei density, $n$(H), and the strength of the incident FUV radiation field, $G$, normalized to the Habing Field in units of $G_{\rm 0} = 1.6 \times 10^{-3}$ erg cm$^{-2}$ s$^{-1}$. In a more realistic geometry of spherical clouds, the optically thin emission would be detected from both the front and back sides of the cloud, whereas the optically thick emission would be detected only from the front side (\citealt{yang2019}). We therefore divide the luminosities of optically thin emission by a factor of two. We also assume that the \cii\ contribution from the PDR is 70\%\ (\S \ref{subsubsec:mol_mass_CII}). We adopt the line ratios before the correction of the CMB and discuss its impact later.

In Figure \ref{fig:pdr_model}, the overlapped region of the four luminosity ratios is log($n$(H)/cm$^{-3}$) $\sim1-5$ with a moderate FUV radiation field $\sim 10^{2}-10^{3}\ G_{\rm 0}$. The high \lcii/$L_{\rm CO(6-5)}$ and \lcii/$L_{\rm CO(7-6)}$ ratios exclude the possibility of log($n$(H)/cm$^{-3}$)$>5$. The strength of the incident FUV radiation field in \name\ is comparable to those in local (U)LIRGs and high-$z$ DSFGs that have $\sim 10^{2}-10^{4} G_{\rm 0}$, as indicated by the grey box in Fig. \ref{fig:pdr_model} (\citealt{hughes2017.cii, wardlow2017}), but lower than that of a $z\sim6$ DSFG, G09.83808, with a FUV radiation field $\sim 10^{4} G_{\rm 0}$(cyan ellipse, \citealt{rybak2020}). The gas density in \name\ is barely constrained, although is lower than that obtained in some $z\sim6-7$ quasar host galaxies (blue square: \citealt{shao2019}). 

The CMB effect makes the intrinsic \lcii/$L_{\rm CO(6-5)}$, \lcii/$L_{\rm CO(7-6)}$, and \lcii/$L_{\rm [CI](2-1)}$ ratios lower because \cii\ is less affected by the CMB compared to CO(6-5), CO(7-6), and \ci. The net effect is that the constraints on $n$(H) and $G$ become weaker.
The high luminosity ratios $> 40$ ($3\sigma$) (Fig. \ref{fig:LFIR_LCII_Lline}) may imply that the nondetection of the lines in \name\ 
could be partly due to its low $n$(H) compared to that of other high-$z$ objects; however, the CMB effect prevents us from obtaining a conclusion. 

\begin{figure}[htbp]
\begin{center}
\includegraphics[width=8cm]{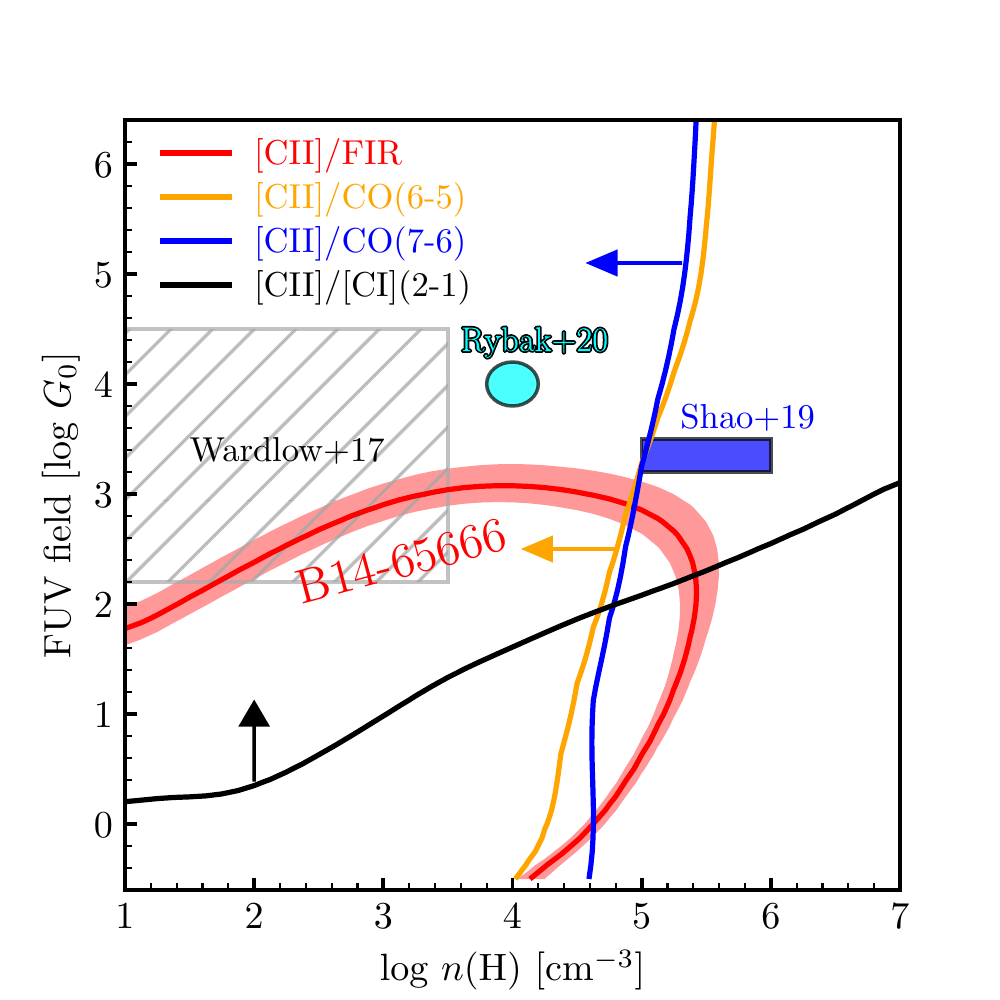}
\end{center}
\caption{
The FUV radiation field, $G$, and the hydrogen gas density, $n$(H), in \name\ as estimated using PDRToolbox (\citealt{pound_wolfire2008}). The red line with a shaded region indicates the parameter space allowed by \lcii/$L_{\rm FIR}$ and its uncertainty. The orange, blue, and black lines with arrows show the permitted ranges of parameters given by the $3\sigma$ lower limits on \lcii/$L_{\rm CO(6-5)}$, \lcii/$L_{\rm CO(7-6)}$, and \lcii/$L_{\rm [CI](2-1)}$, respectively. The allowed parameter space of \name\ corresponds to the red line and shaded region left of the blue and yellow lines. \name\ has log($n$(H)/cm$^{-3}$)$\sim1-5$ and $\sim 10^{2} - 10^{3} G_{\rm 0}$. The results of other DSFGs at $z=1-5$ (\citealt{wardlow2017}), a DSFG at $z=6.0$ (\citealt{rybak2020}), and three IR-bright quasar host galaxies at $z\sim6$ (\citealt{shao2019}) are also shown. 
\label{fig:pdr_model}}
\end{figure}

\subsection{Gas fractions and depletion timescales}\label{subsec:fgas_taudep}

\begin{figure*}[]
\begin{center}
\hspace*{-1cm}
\hspace{+1cm}
\includegraphics[width=9cm]{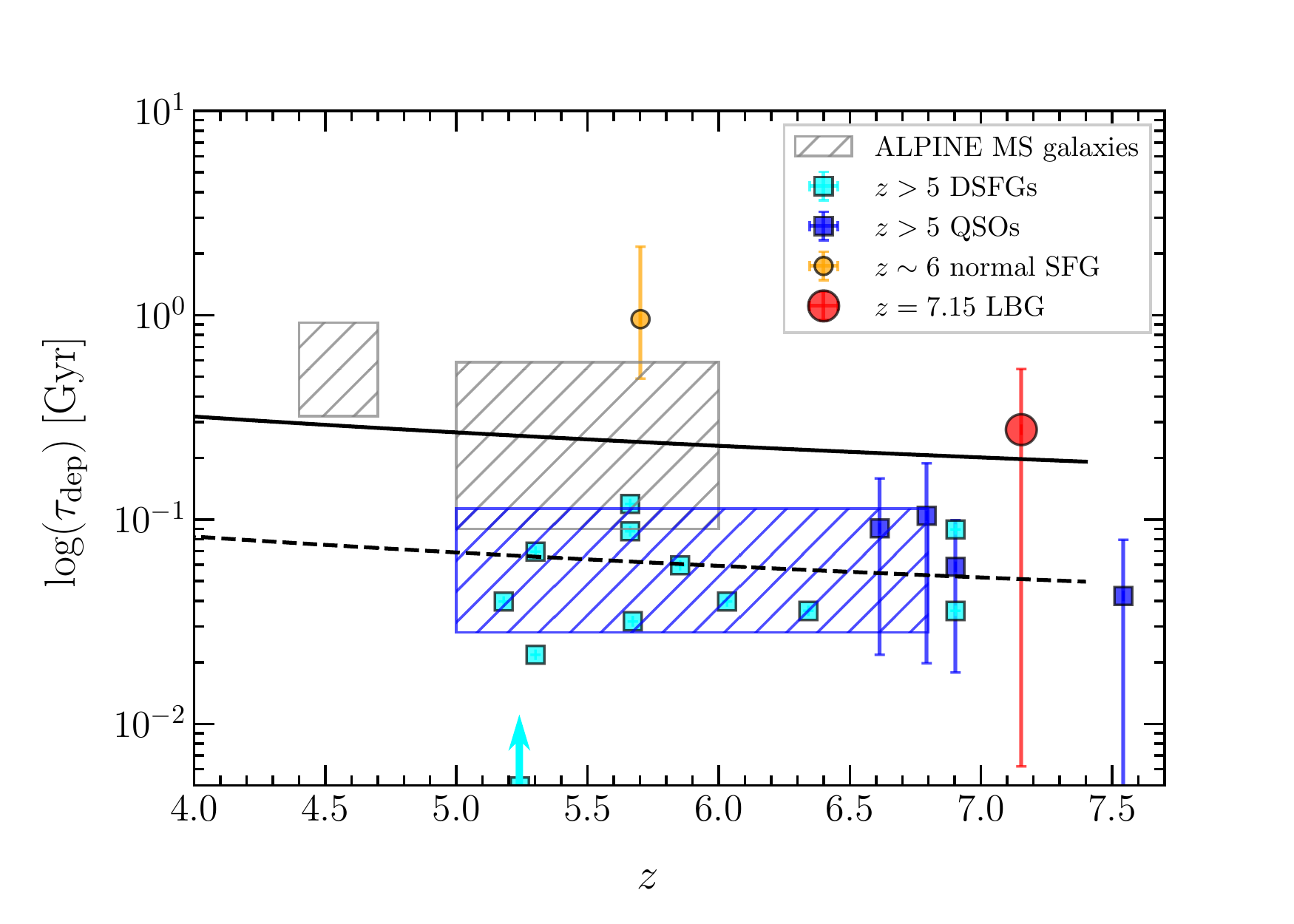}
\hspace{-1cm}
\includegraphics[width=9cm]{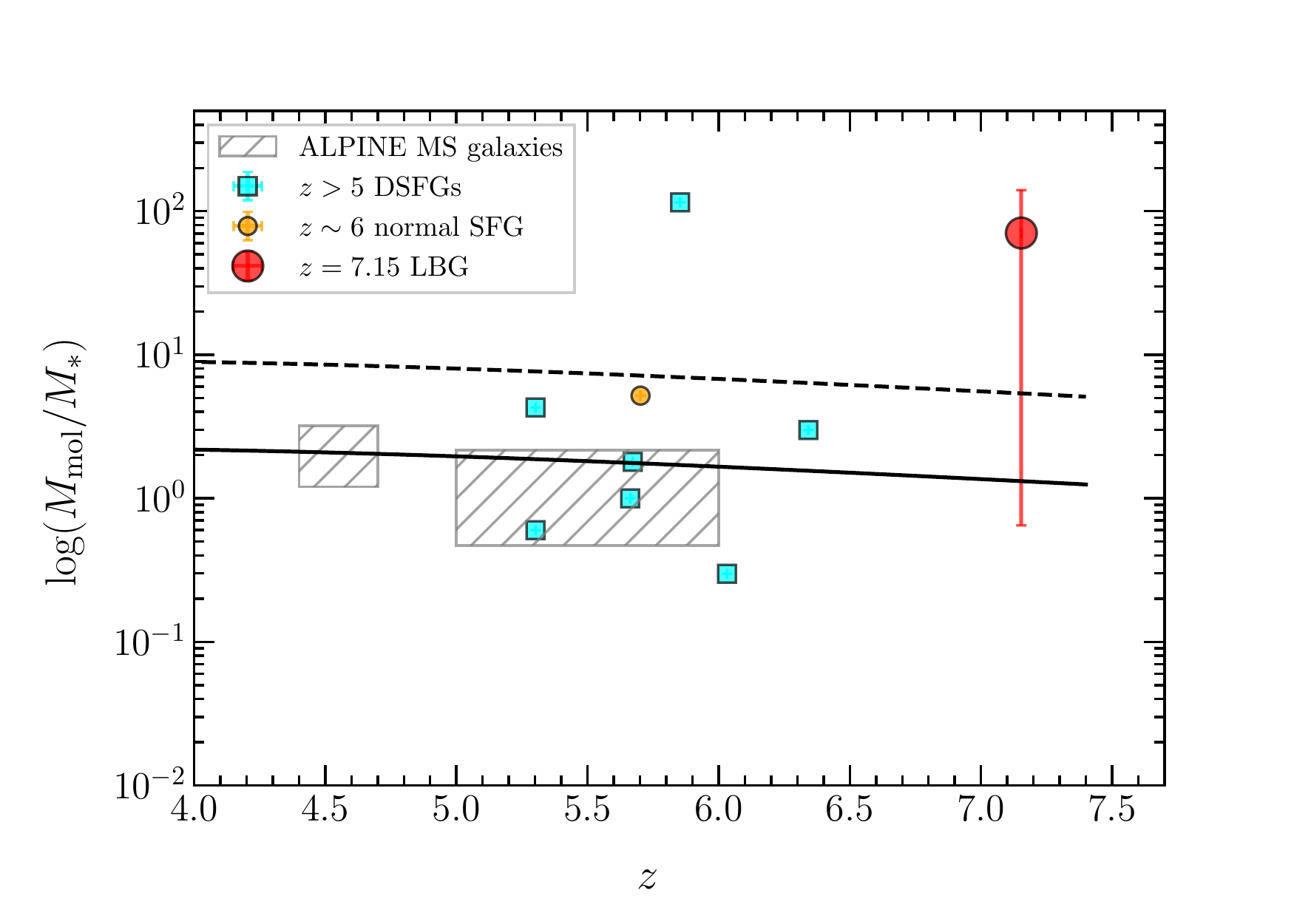}
\end{center}
\vspace*{-0.5cm}
\caption{
Evolution of the gas depletion time ($\tau_{\rm dep}$) and gas fraction ($M_{\rm mol}/M_{\rm *}$) at $z=4-7.5$. (Left) The $\tau_{\rm dep}$ value of \name, indicated by the red circle, is compared against the values in 12 DSFGs at $z>5$ (cyan squares), 10 quasar host galaxies at $z\sim5-6.5$ (blue box regions), four quasar host galaxies at $z\sim7$ (blue squares), a normal SFG at $z\sim6$ (orange circle), and main-sequence galaxies at $z\sim4.5-6.0$ (grey box regions). The solid and dashed lines indicate the scaling relations derived by \cite{tacconi2020} for the main-sequence (log(SFR/SFR$_{\rm MS}$) = 0) and starburst galaxies (log(SFR/SFR$_{\rm MS}$) $=1.2$) at log($M_{\rm *}/M_{\rm \odot}$) $= 10$, respectively, extrapolated to $z\sim7$. 
(Right) Comparisons of $M_{\rm mol}/M_{\rm *}$ in \name\ with those in high-$z$ objects indicated by the same symbols as in the left panel. 
\label{fig:tdep_mugas}}
\end{figure*}

Despite the large uncertainty in $M_{\rm mol}$ (\S \ref{subsec:mol_mass}), we examine two physical quantities related with the molecular gas. The first one is the gas depletion time, $\tau_{\rm dep} \equiv M_{\rm mol}/{\rm SFR}$. With $M_{\rm mol}$ and SFR $= 200^{+82}_{-38}$ \msun\ yr$^{-1}$ from SED fitting (\citealt{hashimoto2019a}), we obtain $\tau_{\rm dep} \approx 2.5 - 550$ Myr. The second one is the molecular gas-to-stellar mass ratio, $\mu_{\rm gas} \equiv M_{\rm mol}/M_{\rm *}$. With $M_{\rm mol} = (0.05 - 11) \times 10^{10}$ \msun\ and the stellar mass obtained from SED fitting (\citealt{hashimoto2019a}), $M_{\rm *} = 7.7^{+1.0}_{-0.8} \times 10^{8}$ \msun, we obtain $\mu_{\rm mol}\approx0.65 - 140$.

The left panel in Fig. \ref{fig:tdep_mugas} shows a comparison of $\tau_{\rm dep}$ of \name\ with other high-$z$ objects shown in Fig. \ref{fig:pLline_LFIR} when the quantities are available. For the 12 data points of $z>5$ DSFGs, we adopt $\tau_{\rm gas}$ in the literatures if available. If not, we compute them from $M_{\rm mol}$ and SFRs. For the 10 quasar host galaxies in \cite{decarli2022}, the width of the box plot corresponds to their redshift range, whereas the height corresponds to the 84 percentile of $\tau_{\rm gas}$. We also individually plot the $z=7.54$ quasar (\citealt{novak2019}) and three $z\sim7$ quasars (\citealt{venemans2017.co.z6}), where $\tau_{\rm gas}$ values are computed based on the combinations of CO-based $M_{\rm mol}$ and SFR. We also plot normal SFGs on the main-sequence at $4 < z < 6$ (\citealt{dessauges-zavadsky2020}) and an individual value of the $z\sim5.7$ LBG, HZ10 (\citealt{pavesi2019}). From the figure, we find that \name\ is consistent with other high-$z$ sources and the extrapolations of the scaling relation in \cite{tacconi2020} (black lines). 

The right panel in Fig. \ref{fig:tdep_mugas} shows a comparison of $\mu_{\rm gas}$. In $z>5$ DSFGs, the values are taken from the literatures if available. If not, we compute them from the stellar and gas mass estimates. We do not include $z>5$ quasar host galaxies because their stellar masses are not well constrained. Again, \name\ is consistent with other high-$z$ sources and the extrapolations of the scaling relation in \cite{tacconi2020}.

\section{Discussion}\label{sec:discussion}

We obtain $\tau_{\rm dep}$ of 2.5 and 550 Myr in the case of $M_{\rm mol} = $ 0.05 and 11 $\times 10^{10}$ \msun, respectively. In the case of $\tau_{\rm dep} = $ 550 (2.5) Myr, the galaxy will consume the molecular gas as early as $z\approx$ 4.5 (7), if the galaxy is not fueled by further accretion, whose final stellar mass is approximately $1\times 10^{11}$ ($1\times10^{9}$) \msun. This implies that \name\ can evolve into a passive galaxy at $z\gtrsim 4$. To further examine this hypothesis, we compare the volume number density of galaxies like \name\ with that of $z\sim3-4$ passive galaxies. The number density of galaxies like \name\ ($M_{\rm UV} = -22.4$) is $\sim 1\times10^{-6}$ Mpc$^{-3}$ based on the bright-end of the UV luminosity function at $z\sim7$ (\citealt{bowler2017, harikane2022}). The observed number density of $z\sim3-4$ passive galaxies was recently compiled by \cite{valentino2020.passive}; it is $\mathcal{O}(10^{-6})-\mathcal{O}(10^{-5})$ Mpc$^{-3}$ at $M_{\rm *}\geq4\times 10^{10}$ \msun. These authors have also derived the number density by analyzing the {\it Illustris} TNG cosmological simulation public release data (e.g. \citealt{springel2018}). In the simulation, the volume number density of $z=3.7$ passive galaxies is estimated to be $\mathcal{O}(10^{-6})$ Mpc$^{-3}$ at $M_{\rm *}\geq4\times 10^{10}$ \msun. A broad agreement in the number densities may support the idea that moderate starburst galaxies at $z>7$, such as \name\, could be ancestors of $z\sim3-4$ passive galaxies (c.f., \citealt{valentino2020.passive}).

\section{Conclusion}\label{sec:conclusion}
We have presented results of ALMA Band 3 observations of CO(6-5), CO(7-6), and \ci(2-1) in \name\ (``Big Three Dragons''). The target was previously detected in \lya, \oiii\ 88 \micron, \cii\ 158 \micron, and the dust continuum in the EoR (\citealt{hashimoto2019a}), and it is one of the brightest LBGs at $z>7$ without gravitational lensing (Table \ref{tab:previous_luminosity}).

\begin{itemize}
\item We do not detect CO(6-5), CO(7-6), and \ci(2-1) (Figs. \ref{fig:maps} and \ref{fig:spectra}). The $3\sigma$ upper limit on the line luminosity is $\approx (2.7-3.0) \times 10^{7}$ \lsun\ [i.e., $(1.7-2.6) \times10^{9}$ K km s$^{-1}$ pc$^{2}$], which is approximately 40 times fainter than the \cii\ 158 \micron\ luminosity before the CMB correction. 

\item By comparing the line luminosity upper limits with $z>5$ sources such as DSFGs and quasar host galaxies, we find that the nondetections are likely due to (1) the insufficient sensitivity of the observations (Fig. \ref{fig:pLline_LFIR}) or (2) possibly low hydrogen gas density in the PDR (Fig. \ref{fig:LFIR_LCII_Lline}), although the uncertainty in the CMB impact makes the interpretation complicated. 
\item We have estimated the molecular gas mass, $M_{\rm mol}$, of \name\ based on five techniques: (1) mid-$J$ CO luminosity, (2) \ci(2-1) luminosity, (3) \cii\ 158 \micron\ luminosity, (4) dust mass and a DGR, and (5) dynamical mass. 
From three methods, namely, \cii, dust mass, and dynamical mass, we obtain $M_{\rm mol} = (0.05-11) \times10^{10}$ \msun, which is consistent with its upper limit inferred from the nondetection of mid-$J$ CO and \ci(2-1) (Table \ref{tab:gas_mass}). 
\item By comparing the observed luminosities to the model predictions of the PDR, we find that \name\ has log($n$(H)/cm$^{-3}$)$\sim1-5$ with a moderate FUV radiation field of $\sim 10^{2} - 10^{3} G_{\rm 0}$. These values are broadly consistent with those obtained in local (U)LIRGs and high-$z$ DSFGs/quasar host galaxies, although the constraints on $n$(H) and $G$ can be weaker if the CMB effect is significant (Fig. \ref{fig:pdr_model}).
\item We estimate a molecular gas-to-stellar mass ratio ($\mu_{\rm gas}$) of $0.65-140$ and a  gas depletion time ($\tau_{\rm dep}$) of $2.5-550$ Myr; these values are consistent with those of other high-$z$ objects and the extrapolations of the scaling relations to $z\sim7$ (Fig. \ref{fig:tdep_mugas}). 
\item 
If the galaxy is not fueled by further accretion, we conjecture that \name\ could be an ancestor of $z\gtrsim 4$ passive galaxies; this is supported by the broad agreement of the number volume density of galaxies like \name\ and $z\sim3-4$ passive galaxies. 
\end{itemize}

\acknowledgments

We thank an anonymous referee for valuable comments that have greatly improved the paper. 
TH was supported by Leading Initiative for Excellent Young Researchers, MEXT, Japan (HJH02007) and by JSPS KAKENHI Grant Numbers (20K22358 and 22H01258). 
AKI, YS, and YF are supported by NAOJ ALMA Scientific Research Grant Numbers 2020-16B. 
AKI acknowledges support from JSPS KAKENHI Grant Number 23H00131. YT acknowledges support from JSPS KAKENHI Grant Number 22H04939. 
KK acknowledges support from the Knut and Alice Wallenberg Foundation.
We thank Yoshito Shimajiri for support in analyzing ALMA data. We appreciate Hajime Fukushima, Shigeki Inoue, Satoshi Kikuta, Masami Ouchi, Hidenobu Yajima, and Atsuhi Yasuda for discussion.
We would like to thank Editage (www.editage.com) for English language editing. 
This paper makes use of the following ALMA data: ADS/JAO.ALMA\#2018.1.01673. ALMA is a partnership of ESO (representing its member states), NSF (USA) and NINS (Japan), together with NRC (Canada), MOST and ASIAA (Taiwan), and KASI (Republic of Korea), in cooperation with the Republic of Chile. The Joint ALMA Observatory is operated by ESO, AUI/NRAO and NAOJ. 
The $HST$ data used in this work are obtained through the data archive at the Space Telescope Science Institute, which is operated by the Association of Universities for Research in Astronomy, Inc. under NASA contract NAS 5-26555.

\vspace{5mm}
\facilities{ALMA}
\clearpage


\begin{thebibliography}{}
\expandafter\ifx\csname natexlab\endcsname\relax\def\natexlab#1{#1}\fi
\providecommand{\url}[1]{\href{#1}{#1}}
\providecommand{\dodoi}[1]{doi:~\href{http://doi.org/#1}{\nolinkurl{#1}}}
\providecommand{\doeprint}[1]{\href{http://ascl.net/#1}{\nolinkurl{http://ascl.net/#1}}}
\providecommand{\doarXiv}[1]{\href{https://arxiv.org/abs/#1}{\nolinkurl{https://arxiv.org/abs/#1}}}

\end{thebibliography}


\begin{thebibliography}{}
\expandafter\ifx\csname natexlab\endcsname\relax\def\natexlab#1{#1}\fi
\providecommand{\url}[1]{\href{#1}{#1}}
\providecommand{\dodoi}[1]{doi:~\href{http://doi.org/#1}{\nolinkurl{#1}}}
\providecommand{\doeprint}[1]{\href{http://ascl.net/#1}{\nolinkurl{http://ascl.net/#1}}}
\providecommand{\doarXiv}[1]{\href{https://arxiv.org/abs/#1}{\nolinkurl{https://arxiv.org/abs/#1}}}

\bibitem[{{Apostolovski} {et~al.}(2019){Apostolovski}, {Aravena}, {Anguita},
  {Spilker}, {Wei{\ss}}, {B{\'e}thermin}, {Chapman}, {Chen}, {Cunningham}, {De
  Breuck}, {Dong}, {Hayward}, {Hezaveh}, {Jarugula}, {Litke}, {Ma}, {Marrone},
  {Narayanan}, {Reuter}, {Rotermund}, \& {Vieira}}]{apostolovski2019}
{Apostolovski}, Y., {Aravena}, M., {Anguita}, T., {et~al.} 2019, \aap, 628,
  A23, \dodoi{10.1051/0004-6361/201935308}

\bibitem[{{Asboth} {et~al.}(2016){Asboth}, {Conley}, {Sayers}, {B{\'e}thermin},
  {Chapman}, {Clements}, {Cooray}, {Dannerbauer}, {Farrah}, {Glenn}, {Golwala},
  {Halpern}, {Ibar}, {Ivison}, {Maloney}, {Marques-Chaves}, {Martinez-Navajas},
  {Oliver}, {P{\'e}rez-Fournon}, {Riechers}, {Rowan-Robinson}, {Scott},
  {Siegel}, {Vieira}, {Viero}, {Wang}, {Wardlow}, \& {Wheeler}}]{asboth2016}
{Asboth}, V., {Conley}, A., {Sayers}, J., {et~al.} 2016, \mnras, 462, 1989,
  \dodoi{10.1093/mnras/stw1769}

\bibitem[{{Bolatto} {et~al.}(2013){Bolatto}, {Wolfire}, \&
  {Leroy}}]{bolatto2013}
{Bolatto}, A.~D., {Wolfire}, M., \& {Leroy}, A.~K. 2013, \araa, 51, 207,
  \dodoi{10.1146/annurev-astro-082812-140944}

\bibitem[{{Bouwens} {et~al.}(2021){Bouwens}, {Oesch}, {Stefanon},
  {Illingworth}, {Labb{\'e}}, {Reddy}, {Atek}, {Montes}, {Naidu},
  {Nanayakkara}, {Nelson}, \& {Wilkins}}]{bouwens2021}
{Bouwens}, R.~J., {Oesch}, P.~A., {Stefanon}, M., {et~al.} 2021, \aj, 162, 47,
  \dodoi{10.3847/1538-3881/abf83e}

\bibitem[{{Bowler} {et~al.}(2018){Bowler}, {Bourne}, {Dunlop}, {McLure}, \&
  {McLeod}}]{bowler2018}
{Bowler}, R.~A.~A., {Bourne}, N., {Dunlop}, J.~S., {McLure}, R.~M., \&
  {McLeod}, D.~J. 2018, ArXiv e-prints.
\newblock \doarXiv{1802.05720}

\bibitem[{{Bowler} {et~al.}(2022){Bowler}, {Cullen}, {McLure}, {Dunlop}, \&
  {Avison}}]{bowler2022}
{Bowler}, R.~A.~A., {Cullen}, F., {McLure}, R.~J., {Dunlop}, J.~S., \&
  {Avison}, A. 2022, \mnras, 510, 5088, \dodoi{10.1093/mnras/stab3744}

\bibitem[{{Bowler} {et~al.}(2017){Bowler}, {Dunlop}, {McLure}, \&
  {McLeod}}]{bowler2017}
{Bowler}, R.~A.~A., {Dunlop}, J.~S., {McLure}, R.~J., \& {McLeod}, D.~J. 2017,
  \mnras, 466, 3612, \dodoi{10.1093/mnras/stw3296}

\bibitem[{{Bowler} {et~al.}(2014){Bowler}, {Dunlop}, {McLure}, {Rogers},
  {McCracken}, {Milvang-Jensen}, {Furusawa}, {Fynbo}, {Taniguchi}, {Afonso},
  {Bremer}, \& {Le F{\`e}vre}}]{bowler2014}
{Bowler}, R.~A.~A., {Dunlop}, J.~S., {McLure}, R.~J., {et~al.} 2014, \mnras,
  440, 2810, \dodoi{10.1093/mnras/stu449}

\bibitem[{{Carilli} \& {Walter}(2013)}]{carilli_walter2013}
{Carilli}, C.~L., \& {Walter}, F. 2013, \araa, 51, 105,
  \dodoi{10.1146/annurev-astro-082812-140953}

\bibitem[{{Casey} {et~al.}(2019){Casey}, {Zavala}, {Aravena}, {B{\'e}thermin},
  {Caputi}, {Champagne}, {Clements}, {da Cunha}, {Drew}, {Finkelstein},
  {Hayward}, {Kartaltepe}, {Knudsen}, {Koekemoer}, {Magdis}, {Man}, {Manning},
  {Scoville}, {Sheth}, {Spilker}, {Staguhn}, {Talia}, {Taniguchi}, {Toft},
  {Treister}, \& {Yun}}]{casey2019}
{Casey}, C.~M., {Zavala}, J.~A., {Aravena}, M., {et~al.} 2019, \apj, 887, 55,
  \dodoi{10.3847/1538-4357/ab52ff}

\bibitem[{{Cicone} {et~al.}(2021){Cicone}, {Mainieri}, {Circosta}, {Kakkad},
  {Vietri}, {Perna}, {Bischetti}, {Carniani}, {Cresci}, {Harrison}, {Mannucci},
  {Marconi}, {Piconcelli}, {Puglisi}, {Scholtz}, {Vignali}, {Zamorani},
  {Zappacosta}, \& {Arrigoni Battaia}}]{cicone2021}
{Cicone}, C., {Mainieri}, V., {Circosta}, C., {et~al.} 2021, \aap, 654, L8,
  \dodoi{10.1051/0004-6361/202141611}

\bibitem[{{Combes} {et~al.}(1999){Combes}, {Maoli}, \& {Omont}}]{combes1999}
{Combes}, F., {Maoli}, R., \& {Omont}, A. 1999, \aap, 345, 369.
\newblock \doarXiv{astro-ph/9902286}

\bibitem[{{Combes} {et~al.}(2012){Combes}, {Rex}, {Rawle}, {Egami}, {Boone},
  {Smail}, {Richard}, {Ivison}, {Gurwell}, {Casey}, {Omont}, {Berciano Alba},
  {Dessauges-Zavadsky}, {Edge}, {Fazio}, {Kneib}, {Okabe}, {Pell{\'o}},
  {P{\'e}rez-Gonz{\'a}lez}, {Schaerer}, {Smith}, {Swinbank}, \& {van der
  Werf}}]{combes2012}
{Combes}, F., {Rex}, M., {Rawle}, T.~D., {et~al.} 2012, \aap, 538, L4,
  \dodoi{10.1051/0004-6361/201118750}

\bibitem[{{Cormier} {et~al.}(2019){Cormier}, {Abel}, {Hony}, {Lebouteiller},
  {Madden}, {Polles}, {Galliano}, {De Looze}, {Galametz}, \&
  {Lambert-Huyghe}}]{cormier2019}
{Cormier}, D., {Abel}, N.~P., {Hony}, S., {et~al.} 2019, \aap, 626, A23,
  \dodoi{10.1051/0004-6361/201834457}

\bibitem[{{Croxall} {et~al.}(2017){Croxall}, {Smith}, {Pellegrini}, {Groves},
  {Bolatto}, {Herrera-Camus}, {Sandstrom}, {Draine}, {Wolfire}, {Armus},
  {Boquien}, {Brandl}, {Dale}, {Galametz}, {Hunt}, {Kennicutt}, {Kreckel},
  {Rigopoulou}, {van der Werf}, \& {Wilson}}]{croxall2017}
{Croxall}, K.~V., {Smith}, J.~D., {Pellegrini}, E., {et~al.} 2017, \apj, 845,
  96, \dodoi{10.3847/1538-4357/aa8035}

\bibitem[{{da Cunha} {et~al.}(2013){da Cunha}, {Groves}, {Walter}, {Decarli},
  {Weiss}, {Bertoldi}, {Carilli}, {Daddi}, {Elbaz}, {Ivison}, {Maiolino},
  {Riechers}, {Rix}, {Sargent}, \& {Smail}}]{da_cunha2013}
{da Cunha}, E., {Groves}, B., {Walter}, F., {et~al.} 2013, \apj, 766, 13,
  \dodoi{10.1088/0004-637X/766/1/13}

\bibitem[{{Decarli} {et~al.}(2022){Decarli}, {Pensabene}, {Venemans}, {Walter},
  {Ba{\~n}ados}, {Bertoldi}, {Carilli}, {Cox}, {Fan}, {Farina}, {Ferkinhoff},
  {Groves}, {Li}, {Mazzucchelli}, {Neri}, {Riechers}, {Uzgil}, {Wang}, {Wang},
  {Weiss}, {Winters}, \& {Yang}}]{decarli2022}
{Decarli}, R., {Pensabene}, A., {Venemans}, B., {et~al.} 2022, \aap, 662, A60,
  \dodoi{10.1051/0004-6361/202142871}

\bibitem[{{Dessauges-Zavadsky} {et~al.}(2020){Dessauges-Zavadsky}, {Ginolfi},
  {Pozzi}, {B{\'e}thermin}, {Le F{\`e}vre}, {Fujimoto}, {Silverman}, {Jones},
  {Vallini}, {Schaerer}, {Faisst}, {Khusanova}, {Fudamoto}, {Cassata},
  {Loiacono}, {Capak}, {Yan}, {Amorin}, {Bardelli}, {Boquien}, {Cimatti},
  {Gruppioni}, {Hathi}, {Ibar}, {Koekemoer}, {Lemaux}, {Narayanan}, {Oesch},
  {Rodighiero}, {Romano}, {Talia}, {Toft}, {Vergani}, {Zamorani}, \&
  {Zucca}}]{dessauges-zavadsky2020}
{Dessauges-Zavadsky}, M., {Ginolfi}, M., {Pozzi}, F., {et~al.} 2020, \aap, 643,
  A5, \dodoi{10.1051/0004-6361/202038231}

\bibitem[{{D'Odorico} {et~al.}(2018){D'Odorico}, {Feruglio}, {Ferrara},
  {Gallerani}, {Pallottini}, {Carniani}, {Maiolino}, {Cristiani}, {Marconi},
  {Piconcelli}, \& {Fiore}}]{dodorico2018}
{D'Odorico}, V., {Feruglio}, C., {Ferrara}, A., {et~al.} 2018, \apjl, 863, L29,
  \dodoi{10.3847/2041-8213/aad7b7}

\bibitem[{{Furusawa} {et~al.}(2016){Furusawa}, {Kashikawa}, {Kobayashi},
  {Dunlop}, {Shimasaku}, {Takata}, {Sekiguchi}, {Naito}, {Furusawa}, {Ouchi},
  {Nakata}, {Yasuda}, {Okura}, {Taniguchi}, {Yamada}, {Kajisawa}, {Fynbo}, \&
  {Le F{\`e}vre}}]{furusawa2016}
{Furusawa}, H., {Kashikawa}, N., {Kobayashi}, M. A.~R., {et~al.} 2016, \apj,
  822, 46, \dodoi{10.3847/0004-637X/822/1/46}

\bibitem[{{Glover} {et~al.}(2015){Glover}, {Clark}, {Micic}, \&
  {Molina}}]{glover2015}
{Glover}, S. C.~O., {Clark}, P.~C., {Micic}, M., \& {Molina}, F. 2015, \mnras,
  448, 1607, \dodoi{10.1093/mnras/stu2699}

\bibitem[{{Goldsmith}(2001)}]{goldsmith2001}
{Goldsmith}, P.~F. 2001, \apj, 557, 736, \dodoi{10.1086/322255}

\bibitem[{{Habing}(1968)}]{habing1968}
{Habing}, H.~J. 1968, \bain, 19, 421

\bibitem[{{Harikane} {et~al.}(2022){Harikane}, {Ono}, {Ouchi}, {Liu},
  {Sawicki}, {Shibuya}, {Behroozi}, {He}, {Shimasaku}, {Arnouts}, {Coupon},
  {Fujimoto}, {Gwyn}, {Huang}, {Inoue}, {Kashikawa}, {Komiyama}, {Matsuoka}, \&
  {Willott}}]{harikane2022}
{Harikane}, Y., {Ono}, Y., {Ouchi}, M., {et~al.} 2022, \apjs, 259, 20,
  \dodoi{10.3847/1538-4365/ac3dfc}

\bibitem[{{Hashimoto} {et~al.}(2019){Hashimoto}, {Inoue}, {Mawatari}, {Tamura},
  {Matsuo}, {Furusawa}, {Harikane}, {Shibuya}, {Knudsen}, {Kohno}, {Ono},
  {Zackrisson}, {Okamoto}, {Kashikawa}, {Oesch}, {Ouchi}, {Ota}, {Shimizu},
  {Taniguchi}, {Umehata}, \& {Watson}}]{hashimoto2019a}
{Hashimoto}, T., {Inoue}, A.~K., {Mawatari}, K., {et~al.} 2019, \pasj, 71, 71,
  \dodoi{10.1093/pasj/psz049}

\bibitem[{{Heintz} \& {Watson}(2020)}]{heintz&watson2020}
{Heintz}, K.~E., \& {Watson}, D. 2020, \apjl, 889, L7,
  \dodoi{10.3847/2041-8213/ab6733}

\bibitem[{{Hopkins} {et~al.}(2008){Hopkins}, {Hernquist}, {Cox}, \&
  {Kere{\v{s}}}}]{hopkins2008}
{Hopkins}, P.~F., {Hernquist}, L., {Cox}, T.~J., \& {Kere{\v{s}}}, D. 2008,
  \apjs, 175, 356, \dodoi{10.1086/524362}

\bibitem[{{Hughes} {et~al.}(2017){Hughes}, {Ibar}, {Villanueva}, {Aravena},
  {Baes}, {Bourne}, {Cooray}, {Dunne}, {Dye}, {Eales}, {Furlanetto},
  {Herrera-Camus}, {Ivison}, {van Kampen}, {Lara-L{\'o}pez}, {Maddox},
  {Micha{\l}owski}, {Smith}, {Valiante}, {van der Werf}, \&
  {Xue}}]{hughes2017.cii}
{Hughes}, T.~M., {Ibar}, E., {Villanueva}, V., {et~al.} 2017, \aap, 602, A49,
  \dodoi{10.1051/0004-6361/201629588}

\bibitem[{{Jarugula} {et~al.}(2021){Jarugula}, {Vieira}, {Weiss}, {Spilker},
  {Aravena}, {Archipley}, {B{\'e}thermin}, {Chapman}, {Dong}, {Greve},
  {Harrington}, {Hayward}, {Hezaveh}, {Hill}, {Litke}, {Malkan}, {Marrone},
  {Narayanan}, {Phadke}, {Reuter}, \& {Rotermund}}]{jarugula2021}
{Jarugula}, S., {Vieira}, J.~D., {Weiss}, A., {et~al.} 2021, \apj, 921, 97,
  \dodoi{10.3847/1538-4357/ac21db}

\bibitem[{{Jiao} {et~al.}(2019){Jiao}, {Zhao}, {Lu}, {Gao}, {Salak}, {Zhu},
  {Zhang}, {Jiang}, \& {Tan}}]{jiao2019}
{Jiao}, Q., {Zhao}, Y., {Lu}, N., {et~al.} 2019, \apj, 880, 133,
  \dodoi{10.3847/1538-4357/ab29ed}

\bibitem[{{Jin} {et~al.}(2019){Jin}, {Daddi}, {Magdis}, {Liu}, {Schinnerer},
  {Papadopoulos}, {Gu}, {Gao}, \& {Calabr{\`o}}}]{jin2019}
{Jin}, S., {Daddi}, E., {Magdis}, G.~E., {et~al.} 2019, \apj, 887, 144,
  \dodoi{10.3847/1538-4357/ab55d6}

\bibitem[{{Kaasinen} {et~al.}(2020){Kaasinen}, {Walter}, {Novak}, {Neeleman},
  {Smail}, {Boogaard}, {Cunha}, {Weiss}, {Liu}, {Decarli}, {Popping},
  {Diaz-Santos}, {Cort{\'e}s}, {Aravena}, {Werf}, {Riechers}, {Inami}, {Hodge},
  {Rix}, \& {Cox}}]{kaasinen2020}
{Kaasinen}, M., {Walter}, F., {Novak}, M., {et~al.} 2020, \apj, 899, 37,
  \dodoi{10.3847/1538-4357/aba438}

\bibitem[{{Kamenetzky} {et~al.}(2016){Kamenetzky}, {Rangwala}, {Glenn},
  {Maloney}, \& {Conley}}]{kamenetzky2016}
{Kamenetzky}, J., {Rangwala}, N., {Glenn}, J., {Maloney}, P.~R., \& {Conley},
  A. 2016, \apj, 829, 93, \dodoi{10.3847/0004-637X/829/2/93}

\bibitem[{{Kaufman} {et~al.}(2006){Kaufman}, {Wolfire}, \&
  {Hollenbach}}]{kaufman2006}
{Kaufman}, M.~J., {Wolfire}, M.~G., \& {Hollenbach}, D.~J. 2006, \apj, 644,
  283, \dodoi{10.1086/503596}

\bibitem[{{Kohandel} {et~al.}(2019){Kohandel}, {Pallottini}, {Ferrara},
  {Zanella}, {Behrens}, {Carniani}, {Gallerani}, \& {Vallini}}]{kohandel2019}
{Kohandel}, M., {Pallottini}, A., {Ferrara}, A., {et~al.} 2019, \mnras, 487,
  3007, \dodoi{10.1093/mnras/stz1486}

\bibitem[{{Komatsu} {et~al.}(2011){Komatsu}, {Smith}, {Dunkley}, {Bennett},
  {Gold}, {Hinshaw}, {Jarosik}, {Larson}, {Nolta}, {Page}, {Spergel},
  {Halpern}, {Hill}, {Kogut}, {Limon}, {Meyer}, {Odegard}, {Tucker}, {Weiland},
  {Wollack}, \& {Wright}}]{komatsu2011}
{Komatsu}, E., {Smith}, K.~M., {Dunkley}, J., {et~al.} 2011, \apjs, 192, 18,
  \dodoi{10.1088/0067-0049/192/2/18}

\bibitem[{{Lagache} {et~al.}(2018){Lagache}, {Cousin}, \&
  {Chatzikos}}]{lagache2018}
{Lagache}, G., {Cousin}, M., \& {Chatzikos}, M. 2018, \aap, 609, A130,
  \dodoi{10.1051/0004-6361/201732019}

\bibitem[{{Leroy} {et~al.}(2011){Leroy}, {Bolatto}, {Gordon}, {Sandstrom},
  {Gratier}, {Rosolowsky}, {Engelbracht}, {Mizuno}, {Corbelli}, {Fukui}, \&
  {Kawamura}}]{leroy2011}
{Leroy}, A.~K., {Bolatto}, A., {Gordon}, K., {et~al.} 2011, \apj, 737, 12,
  \dodoi{10.1088/0004-637X/737/1/12}

\bibitem[{{Li} {et~al.}(2020){Li}, {Wang}, {Riechers}, {Walter}, {Decarli},
  {Venamans}, {Neri}, {Shao}, {Fan}, {Gao}, {Carilli}, {Omont}, {Cox},
  {Menten}, {Wagg}, {Bertoldi}, \& {Narayanan}}]{li2020.co}
{Li}, J., {Wang}, R., {Riechers}, D., {et~al.} 2020, \apj, 889, 162,
  \dodoi{10.3847/1538-4357/ab65fa}

\bibitem[{{Li} {et~al.}(2019){Li}, {Narayanan}, \& {Dav{\'e}}}]{li2019}
{Li}, Q., {Narayanan}, D., \& {Dav{\'e}}, R. 2019, \mnras, 490, 1425,
  \dodoi{10.1093/mnras/stz2684}

\bibitem[{{Madden} {et~al.}(2020){Madden}, {Cormier}, {Hony}, {Lebouteiller},
  {Abel}, {Galametz}, {De Looze}, {Chevance}, {Polles}, {Lee}, {Galliano},
  {Lambert-Huyghe}, {Hu}, \& {Ramambason}}]{madden2020}
{Madden}, S.~C., {Cormier}, D., {Hony}, S., {et~al.} 2020, \aap, 643, A141,
  \dodoi{10.1051/0004-6361/202038860}

\bibitem[{{Magdis} {et~al.}(2012){Magdis}, {Daddi}, {B{\'e}thermin}, {Sargent},
  {Elbaz}, {Pannella}, {Dickinson}, {Dannerbauer}, {da Cunha}, {Walter},
  {Rigopoulou}, {Charmandaris}, {Hwang}, \& {Kartaltepe}}]{magdis2012}
{Magdis}, G.~E., {Daddi}, E., {B{\'e}thermin}, M., {et~al.} 2012, \apj, 760, 6,
  \dodoi{10.1088/0004-637X/760/1/6}

\bibitem[{{McCracken} {et~al.}(2012){McCracken}, {Milvang-Jensen}, {Dunlop},
  {Franx}, {Fynbo}, {Le F{\`e}vre}, {Holt}, {Caputi}, {Goranova}, {Buitrago},
  {Emerson}, {Freudling}, {Hudelot}, {L{\'o}pez-Sanjuan}, {Magnard}, {Mellier},
  {M{\o}ller}, {Nilsson}, {Sutherland}, {Tasca}, \& {Zabl}}]{mccracken2012}
{McCracken}, H.~J., {Milvang-Jensen}, B., {Dunlop}, J., {et~al.} 2012, \aap,
  544, A156, \dodoi{10.1051/0004-6361/201219507}

\bibitem[{{McMullin} {et~al.}(2007){McMullin}, {Waters}, {Schiebel}, {Young},
  \& {Golap}}]{McMullin2007}
{McMullin}, J.~P., {Waters}, B., {Schiebel}, D., {Young}, W., \& {Golap}, K.
  2007, in Astronomical Society of the Pacific Conference Series, Vol. 376,
  Astronomical Data Analysis Software and Systems XVI, ed. R.~A. {Shaw},
  F.~{Hill}, \& D.~J. {Bell}, 127

\bibitem[{{Meijerink} {et~al.}(2007){Meijerink}, {Spaans}, \&
  {Israel}}]{meijerink2007}
{Meijerink}, R., {Spaans}, M., \& {Israel}, F.~P. 2007, \aap, 461, 793,
  \dodoi{10.1051/0004-6361:20066130}

\bibitem[{{Narayanan} \& {Krumholz}(2014)}]{narayanan2014}
{Narayanan}, D., \& {Krumholz}, M.~R. 2014, \mnras, 442, 1411,
  \dodoi{10.1093/mnras/stu834}

\bibitem[{{Narayanan} {et~al.}(2012){Narayanan}, {Krumholz}, {Ostriker}, \&
  {Hernquist}}]{narayanan2012}
{Narayanan}, D., {Krumholz}, M.~R., {Ostriker}, E.~C., \& {Hernquist}, L. 2012,
  \mnras, 421, 3127, \dodoi{10.1111/j.1365-2966.2012.20536.x}

\bibitem[{{Novak} {et~al.}(2019){Novak}, {Ba{\~n}ados}, {Decarli}, {Walter},
  {Venemans}, {Neeleman}, {Farina}, {Mazzucchelli}, {Carilli}, {Fan}, {Rix}, \&
  {Wang}}]{novak2019}
{Novak}, M., {Ba{\~n}ados}, E., {Decarli}, R., {et~al.} 2019, \apj, 881, 63,
  \dodoi{10.3847/1538-4357/ab2beb}

\bibitem[{{Obreschkow} {et~al.}(2009){Obreschkow}, {Heywood}, {Kl{\"o}ckner},
  \& {Rawlings}}]{obreschkow2009}
{Obreschkow}, D., {Heywood}, I., {Kl{\"o}ckner}, H.~R., \& {Rawlings}, S. 2009,
  \apj, 702, 1321, \dodoi{10.1088/0004-637X/702/2/1321}

\bibitem[{{Offner} {et~al.}(2014){Offner}, {Bisbas}, {Bell}, \&
  {Viti}}]{offner2014}
{Offner}, S.~S.~R., {Bisbas}, T.~G., {Bell}, T.~A., \& {Viti}, S. 2014, \mnras,
  440, L81, \dodoi{10.1093/mnrasl/slu013}

\bibitem[{{Oke} \& {Gunn}(1983)}]{oke1983}
{Oke}, J.~B., \& {Gunn}, J.~E. 1983, \apj, 266, 713, \dodoi{10.1086/160817}

\bibitem[{{Papadopoulos} {et~al.}(2018){Papadopoulos}, {Bisbas}, \&
  {Zhang}}]{papadopoulos2018}
{Papadopoulos}, P.~P., {Bisbas}, T.~G., \& {Zhang}, Z.-Y. 2018, \mnras, 478,
  1716, \dodoi{10.1093/mnras/sty1077}

\bibitem[{{Papadopoulos} {et~al.}(2000){Papadopoulos}, {R{\"o}ttgering}, {van
  der Werf}, {Guilloteau}, {Omont}, {van Breugel}, \&
  {Tilanus}}]{papadopoulos2000}
{Papadopoulos}, P.~P., {R{\"o}ttgering}, H.~J.~A., {van der Werf}, P.~P.,
  {et~al.} 2000, \apj, 528, 626, \dodoi{10.1086/308215}

\bibitem[{{Pavesi} {et~al.}(2019){Pavesi}, {Riechers}, {Faisst}, {Stacey}, \&
  {Capak}}]{pavesi2019}
{Pavesi}, R., {Riechers}, D.~A., {Faisst}, A.~L., {Stacey}, G.~J., \& {Capak},
  P.~L. 2019, \apj, 882, 168, \dodoi{10.3847/1538-4357/ab3a46}

\bibitem[{{Pavesi} {et~al.}(2018){Pavesi}, {Sharon}, {Riechers}, {Hodge},
  {Decarli}, {Walter}, {Carilli}, {Daddi}, {Smail}, {Dickinson}, {Ivison},
  {Sargent}, {da Cunha}, {Aravena}, {Darling}, {Smol{\v{c}}i{\'c}}, {Scoville},
  {Capak}, \& {Wagg}}]{pavesi2018}
{Pavesi}, R., {Sharon}, C.~E., {Riechers}, D.~A., {et~al.} 2018, \apj, 864, 49,
  \dodoi{10.3847/1538-4357/aacb79}

\bibitem[{{Pound} \& {Wolfire}(2008)}]{pound_wolfire2008}
{Pound}, M.~W., \& {Wolfire}, M.~G. 2008, in Astronomical Society of the
  Pacific Conference Series, Vol. 394, Astronomical Data Analysis Software and
  Systems XVII, ed. R.~W. {Argyle}, P.~S. {Bunclark}, \& J.~R. {Lewis}, 654

\bibitem[{{Rawle} {et~al.}(2014){Rawle}, {Egami}, {Bussmann}, {Gurwell},
  {Ivison}, {Boone}, {Combes}, {Danielson}, {Rex}, {Richard}, {Smail},
  {Swinbank}, {Altieri}, {Blain}, {Clement}, {Dessauges-Zavadsky}, {Edge},
  {Fazio}, {Jones}, {Kneib}, {Omont}, {P{\'e}rez-Gonz{\'a}lez}, {Schaerer},
  {Valtchanov}, {van der Werf}, {Walth}, {Zamojski}, \& {Zemcov}}]{rawle2014}
{Rawle}, T.~D., {Egami}, E., {Bussmann}, R.~S., {et~al.} 2014, \apj, 783, 59,
  \dodoi{10.1088/0004-637X/783/1/59}

\bibitem[{{R{\'e}my-Ruyer} {et~al.}(2014){R{\'e}my-Ruyer}, {Madden},
  {Galliano}, {Galametz}, {Takeuchi}, {Asano}, {Zhukovska}, {Lebouteiller},
  {Cormier}, {Jones}, {Bocchio}, {Baes}, {Bendo}, {Boquien}, {Boselli},
  {DeLooze}, {Doublier-Pritchard}, {Hughes}, {Karczewski}, \&
  {Spinoglio}}]{remy-ruyer2014}
{R{\'e}my-Ruyer}, A., {Madden}, S.~C., {Galliano}, F., {et~al.} 2014, \aap,
  563, A31, \dodoi{10.1051/0004-6361/201322803}

\bibitem[{{Riechers} {et~al.}(2009){Riechers}, {Walter}, {Bertoldi}, {Carilli},
  {Aravena}, {Neri}, {Cox}, {Wei{\ss}}, \& {Menten}}]{riechers2009}
{Riechers}, D.~A., {Walter}, F., {Bertoldi}, F., {et~al.} 2009, \apj, 703,
  1338, \dodoi{10.1088/0004-637X/703/2/1338}

\bibitem[{{Riechers} {et~al.}(2013){Riechers}, {Bradford}, {Clements},
  {Dowell}, {P{\'e}rez-Fournon}, {Ivison}, {Bridge}, {Conley}, {Fu}, {Vieira},
  {Wardlow}, {Calanog}, {Cooray}, {Hurley}, {Neri}, {Kamenetzky}, {Aguirre},
  {Altieri}, {Arumugam}, {Benford}, {B{\'e}thermin}, {Bock}, {Burgarella},
  {Cabrera-Lavers}, {Chapman}, {Cox}, {Dunlop}, {Earle}, {Farrah}, {Ferrero},
  {Franceschini}, {Gavazzi}, {Glenn}, {Solares}, {Gurwell}, {Halpern},
  {Hatziminaoglou}, {Hyde}, {Ibar}, {Kov{\'a}cs}, {Krips}, {Lupu}, {Maloney},
  {Martinez-Navajas}, {Matsuhara}, {Murphy}, {Naylor}, {Nguyen}, {Oliver},
  {Omont}, {Page}, {Petitpas}, {Rangwala}, {Roseboom}, {Scott}, {Smith},
  {Staguhn}, {Streblyanska}, {Thomson}, {Valtchanov}, {Viero}, {Wang},
  {Zemcov}, \& {Zmuidzinas}}]{riechers2013}
{Riechers}, D.~A., {Bradford}, C.~M., {Clements}, D.~L., {et~al.} 2013, \nat,
  496, 329, \dodoi{10.1038/nature12050}

\bibitem[{{Riechers} {et~al.}(2017){Riechers}, {Leung}, {Ivison},
  {P{\'e}rez-Fournon}, {Lewis}, {Marques-Chaves}, {Oteo}, {Clements}, {Cooray},
  {Greenslade}, {Mart{\'\i}nez-Navajas}, {Oliver}, {Rigopoulou}, {Scott}, \&
  {Weiss}}]{riechers2017}
{Riechers}, D.~A., {Leung}, T.~K.~D., {Ivison}, R.~J., {et~al.} 2017, \apj,
  850, 1, \dodoi{10.3847/1538-4357/aa8ccf}

\bibitem[{{Riechers} {et~al.}(2020){Riechers}, {Hodge}, {Pavesi}, {Daddi},
  {Decarli}, {Ivison}, {Sharon}, {Smail}, {Walter}, {Aravena}, {Capak},
  {Carilli}, {Cox}, {Cunha}, {Dannerbauer}, {Dickinson}, {Neri}, \&
  {Wagg}}]{riechers2020}
{Riechers}, D.~A., {Hodge}, J.~A., {Pavesi}, R., {et~al.} 2020, \apj, 895, 81,
  \dodoi{10.3847/1538-4357/ab8c48}

\bibitem[{{Riechers} {et~al.}(2021){Riechers}, {Nayyeri}, {Burgarella},
  {Emonts}, {Clements}, {Cooray}, {Ivison}, {Oliver}, {P{\'e}rez-Fournon},
  {Rigopoulou}, \& {Scott}}]{riechers2021}
{Riechers}, D.~A., {Nayyeri}, H., {Burgarella}, D., {et~al.} 2021, \apj, 907,
  62, \dodoi{10.3847/1538-4357/abcf2e}

\bibitem[{{Rybak} {et~al.}(2020){Rybak}, {Zavala}, {Hodge}, {Casey}, \&
  {Werf}}]{rybak2020}
{Rybak}, M., {Zavala}, J.~A., {Hodge}, J.~A., {Casey}, C.~M., \& {Werf}, P.
  v.~d. 2020, \apjl, 889, L11, \dodoi{10.3847/2041-8213/ab63de}

\bibitem[{{Sakamoto}(1999)}]{sakamoto1999}
{Sakamoto}, S. 1999, \apj, 523, 701, \dodoi{10.1086/307741}

\bibitem[{{Scoville} {et~al.}(2016){Scoville}, {Sheth}, {Aussel}, {Vanden
  Bout}, {Capak}, {Bongiorno}, {Casey}, {Murchikova}, {Koda},
  {{\'A}lvarez-M{\'a}rquez}, {Lee}, {Laigle}, {McCracken}, {Ilbert}, {Pope},
  {Sanders}, {Chu}, {Toft}, {Ivison}, \& {Manohar}}]{scoville2016}
{Scoville}, N., {Sheth}, K., {Aussel}, H., {et~al.} 2016, \apj, 820, 83,
  \dodoi{10.3847/0004-637X/820/2/83}

\bibitem[{{Shao} {et~al.}(2019){Shao}, {Wang}, {Carilli}, {Wagg}, {Walter},
  {Li}, {Fan}, {Jiang}, {Riechers}, {Bertoldi}, {Strauss}, {Cox}, {Omont}, \&
  {Menten}}]{shao2019}
{Shao}, Y., {Wang}, R., {Carilli}, C.~L., {et~al.} 2019, \apj, 876, 99,
  \dodoi{10.3847/1538-4357/ab133d}

\bibitem[{{Shi} {et~al.}(2016){Shi}, {Wang}, {Zhang}, {Gao}, {Hao}, {Xia}, \&
  {Gu}}]{shi2016}
{Shi}, Y., {Wang}, J., {Zhang}, Z.-Y., {et~al.} 2016, Nature Communications, 7,
  13789, \dodoi{10.1038/ncomms13789}

\bibitem[{{Shimajiri} {et~al.}(2013){Shimajiri}, {Sakai}, {Tsukagoshi},
  {Kitamura}, {Momose}, {Saito}, {Oshima}, {Kohno}, \&
  {Kawabe}}]{shimajiri2013}
{Shimajiri}, Y., {Sakai}, T., {Tsukagoshi}, T., {et~al.} 2013, \apjl, 774, L20,
  \dodoi{10.1088/2041-8205/774/2/L20}

\bibitem[{{Solomon} {et~al.}(1992){Solomon}, {Downes}, \&
  {Radford}}]{solomon1992}
{Solomon}, P.~M., {Downes}, D., \& {Radford}, S.~J.~E. 1992, \apjl, 398, L29,
  \dodoi{10.1086/186569}

\bibitem[{{Springel} {et~al.}(2018){Springel}, {Pakmor}, {Pillepich},
  {Weinberger}, {Nelson}, {Hernquist}, {Vogelsberger}, {Genel}, {Torrey},
  {Marinacci}, \& {Naiman}}]{springel2018}
{Springel}, V., {Pakmor}, R., {Pillepich}, A., {et~al.} 2018, \mnras, 475, 676,
  \dodoi{10.1093/mnras/stx3304}

\bibitem[{{Stefan} {et~al.}(2015){Stefan}, {Carilli}, {Wagg}, {Walter},
  {Riechers}, {Bertoldi}, {Green}, {Fan}, {Menten}, \& {Wang}}]{stefan2015}
{Stefan}, I.~I., {Carilli}, C.~L., {Wagg}, J., {et~al.} 2015, \mnras, 451,
  1713, \dodoi{10.1093/mnras/stv1108}

\bibitem[{{Strandet} {et~al.}(2016){Strandet}, {Weiss}, {Vieira}, {de Breuck},
  {Aguirre}, {Aravena}, {Ashby}, {B{\'e}thermin}, {Bradford}, {Carlstrom},
  {Chapman}, {Crawford}, {Everett}, {Fassnacht}, {Furstenau}, {Gonzalez},
  {Greve}, {Gullberg}, {Hezaveh}, {Kamenetzky}, {Litke}, {Ma}, {Malkan},
  {Marrone}, {Menten}, {Murphy}, {Nadolski}, {Rotermund}, {Spilker}, {Stark},
  \& {Welikala}}]{strandet2016}
{Strandet}, M.~L., {Weiss}, A., {Vieira}, J.~D., {et~al.} 2016, \apj, 822, 80,
  \dodoi{10.3847/0004-637X/822/2/80}

\bibitem[{{Strandet} {et~al.}(2017){Strandet}, {Weiss}, {De Breuck}, {Marrone},
  {Vieira}, {Aravena}, {Ashby}, {B{\'e}thermin}, {Bothwell}, {Bradford},
  {Carlstrom}, {Chapman}, {Cunningham}, {Chen}, {Fassnacht}, {Gonzalez},
  {Greve}, {Gullberg}, {Hayward}, {Hezaveh}, {Litke}, {Ma}, {Malkan}, {Menten},
  {Miller}, {Murphy}, {Narayanan}, {Phadke}, {Rotermund}, {Spilker}, \&
  {Sreevani}}]{strandet2017}
{Strandet}, M.~L., {Weiss}, A., {De Breuck}, C., {et~al.} 2017, \apjl, 842,
  L15, \dodoi{10.3847/2041-8213/aa74b0}

\bibitem[{{Sugahara} {et~al.}(2021){Sugahara}, {Inoue}, {Hashimoto},
  {Yamanaka}, {Fujimoto}, {Tamura}, {Matsuo}, {Binggeli}, \&
  {Zackrisson}}]{sugahara2021}
{Sugahara}, Y., {Inoue}, A.~K., {Hashimoto}, T., {et~al.} 2021, \apj, 923, 5,
  \dodoi{10.3847/1538-4357/ac2a36}

\bibitem[{{Tacconi} {et~al.}(2020){Tacconi}, {Genzel}, \&
  {Sternberg}}]{tacconi2020}
{Tacconi}, L.~J., {Genzel}, R., \& {Sternberg}, A. 2020, \araa, 58, 157,
  \dodoi{10.1146/annurev-astro-082812-141034}

\bibitem[{{Tacconi} {et~al.}(2018){Tacconi}, {Genzel}, {Saintonge}, {Combes},
  {Garc{\'\i}a-Burillo}, {Neri}, {Bolatto}, {Contini}, {F{\"o}rster Schreiber},
  {Lilly}, {Lutz}, {Wuyts}, {Accurso}, {Boissier}, {Boone}, {Bouch{\'e}},
  {Bournaud}, {Burkert}, {Carollo}, {Cooper}, {Cox}, {Feruglio}, {Freundlich},
  {Herrera-Camus}, {Juneau}, {Lippa}, {Naab}, {Renzini}, {Salome}, {Sternberg},
  {Tadaki}, {{\"U}bler}, {Walter}, {Weiner}, \& {Weiss}}]{tacconi2018}
{Tacconi}, L.~J., {Genzel}, R., {Saintonge}, A., {et~al.} 2018, \apj, 853, 179,
  \dodoi{10.3847/1538-4357/aaa4b4}

\bibitem[{{Tunnard} \& {Greve}(2016)}]{tunnard_greve2016}
{Tunnard}, R., \& {Greve}, T.~R. 2016, \apj, 819, 161,
  \dodoi{10.3847/0004-637X/819/2/161}

\bibitem[{{Valentino} {et~al.}(2018){Valentino}, {Magdis}, {Daddi}, {Liu},
  {Aravena}, {Bournaud}, {Cibinel}, {Cormier}, {Dickinson}, {Gao}, {Jin},
  {Juneau}, {Kartaltepe}, {Lee}, {Madden}, {Puglisi}, {Sanders}, \&
  {Silverman}}]{valentino2018}
{Valentino}, F., {Magdis}, G.~E., {Daddi}, E., {et~al.} 2018, \apj, 869, 27,
  \dodoi{10.3847/1538-4357/aaeb88}

\bibitem[{{Valentino} {et~al.}(2020){Valentino}, {Tanaka}, {Davidzon}, {Toft},
  {G{\'o}mez-Guijarro}, {Stockmann}, {Onodera}, {Brammer}, {Ceverino},
  {Faisst}, {Gallazzi}, {Hayward}, {Ilbert}, {Kubo}, {Magdis}, {Selsing},
  {Shimakawa}, {Sparre}, {Steinhardt}, {Yabe}, \&
  {Zabl}}]{valentino2020.passive}
{Valentino}, F., {Tanaka}, M., {Davidzon}, I., {et~al.} 2020, \apj, 889, 93,
  \dodoi{10.3847/1538-4357/ab64dc}

\bibitem[{{Vallini} {et~al.}(2015){Vallini}, {Gallerani}, {Ferrara},
  {Pallottini}, \& {Yue}}]{vallini2015}
{Vallini}, L., {Gallerani}, S., {Ferrara}, A., {Pallottini}, A., \& {Yue}, B.
  2015, \apj, 813, 36, \dodoi{10.1088/0004-637X/813/1/36}

\bibitem[{{Vallini} {et~al.}(2019){Vallini}, {Tielens}, {Pallottini},
  {Gallerani}, {Gruppioni}, {Carniani}, {Pozzi}, \& {Talia}}]{vallini2019}
{Vallini}, L., {Tielens}, A.~G.~G.~M., {Pallottini}, A., {et~al.} 2019, \mnras,
  490, 4502, \dodoi{10.1093/mnras/stz2837}

\bibitem[{{Venemans} {et~al.}(2017{\natexlab{a}}){Venemans}, {Walter},
  {Decarli}, {Ba{\~n}ados}, {Hodge}, {Hewett}, {McMahon}, {Mortlock}, \&
  {Simpson}}]{venemans2017.co.z7}
{Venemans}, B.~P., {Walter}, F., {Decarli}, R., {et~al.} 2017{\natexlab{a}},
  \apj, 837, 146, \dodoi{10.3847/1538-4357/aa62ac}

\bibitem[{{Venemans} {et~al.}(2017{\natexlab{b}}){Venemans}, {Walter},
  {Decarli}, {Ferkinhoff}, {Wei{\ss}}, {Findlay}, {McMahon}, {Sutherland}, \&
  {Meijerink}}]{venemans2017.co.z6}
---. 2017{\natexlab{b}}, \apj, 845, 154, \dodoi{10.3847/1538-4357/aa81cb}

\bibitem[{{Vieira} {et~al.}(2022){Vieira}, {Riechers}, {Pavesi}, {Faisst},
  {Schinnerer}, {Scoville}, \& {Stacey}}]{vieira2022}
{Vieira}, D., {Riechers}, D.~A., {Pavesi}, R., {et~al.} 2022, \apj, 925, 174,
  \dodoi{10.3847/1538-4357/ac403a}

\bibitem[{{Vieira} {et~al.}(2013){Vieira}, {Marrone}, {Chapman}, {De Breuck},
  {Hezaveh}, {Wei{\ensuremath{\beta}}}, {Aguirre}, {Aird}, {Aravena}, {Ashby},
  {Bayliss}, {Benson}, {Biggs}, {Bleem}, {Bock}, {Bothwell}, {Bradford},
  {Brodwin}, {Carlstrom}, {Chang}, {Crawford}, {Crites}, {de Haan}, {Dobbs},
  {Fomalont}, {Fassnacht}, {George}, {Gladders}, {Gonzalez}, {Greve},
  {Gullberg}, {Halverson}, {High}, {Holder}, {Holzapfel}, {Hoover}, {Hrubes},
  {Hunter}, {Keisler}, {Lee}, {Leitch}, {Lueker}, {Luong-van}, {Malkan},
  {McIntyre}, {McMahon}, {Mehl}, {Menten}, {Meyer}, {Mocanu}, {Murphy},
  {Natoli}, {Padin}, {Plagge}, {Reichardt}, {Rest}, {Ruel}, {Ruhl}, {Sharon},
  {Schaffer}, {Shaw}, {Shirokoff}, {Spilker}, {Stalder}, {Staniszewski},
  {Stark}, {Story}, {Vanderlinde}, {Welikala}, \& {Williamson}}]{vieira2013}
{Vieira}, J.~D., {Marrone}, D.~P., {Chapman}, S.~C., {et~al.} 2013, \nat, 495,
  344, \dodoi{10.1038/nature12001}

\bibitem[{{Walter} {et~al.}(2003){Walter}, {Bertoldi}, {Carilli}, {Cox}, {Lo},
  {Neri}, {Fan}, {Omont}, {Strauss}, \& {Menten}}]{walter2003}
{Walter}, F., {Bertoldi}, F., {Carilli}, C., {et~al.} 2003, \nat, 424, 406,
  \dodoi{10.1038/nature01821}

\bibitem[{{Wang} {et~al.}(2019){Wang}, {Wang}, {Fan}, {Wu}, {Yang}, {Neri}, \&
  {Yue}}]{feige.wang2019}
{Wang}, F., {Wang}, R., {Fan}, X., {et~al.} 2019, \apj, 880, 2,
  \dodoi{10.3847/1538-4357/ab2717}

\bibitem[{{Wang} {et~al.}(2010){Wang}, {Carilli}, {Neri}, {Riechers}, {Wagg},
  {Walter}, {Bertoldi}, {Menten}, {Omont}, {Cox}, \& {Fan}}]{wang2010}
{Wang}, R., {Carilli}, C.~L., {Neri}, R., {et~al.} 2010, \apj, 714, 699,
  \dodoi{10.1088/0004-637X/714/1/699}

\bibitem[{{Wang} {et~al.}(2011{\natexlab{a}}){Wang}, {Wagg}, {Carilli},
  {Walter}, {Riechers}, {Willott}, {Bertoldi}, {Omont}, {Beelen}, {Cox},
  {Strauss}, {Bergeron}, {Forveille}, {Menten}, \& {Fan}}]{wang2011a}
{Wang}, R., {Wagg}, J., {Carilli}, C.~L., {et~al.} 2011{\natexlab{a}}, \apjl,
  739, L34, \dodoi{10.1088/2041-8205/739/1/L34}

\bibitem[{{Wang} {et~al.}(2011{\natexlab{b}}){Wang}, {Wagg}, {Carilli}, {Neri},
  {Walter}, {Omont}, {Riechers}, {Bertoldi}, {Menten}, {Cox}, {Strauss}, {Fan},
  \& {Jiang}}]{wang2011b}
---. 2011{\natexlab{b}}, \aj, 142, 101, \dodoi{10.1088/0004-6256/142/4/101}

\bibitem[{{Wang} {et~al.}(2013){Wang}, {Wagg}, {Carilli}, {Walter}, {Lentati},
  {Fan}, {Riechers}, {Bertoldi}, {Narayanan}, {Strauss}, {Cox}, {Omont},
  {Menten}, {Knudsen}, {Neri}, \& {Jiang}}]{wang2013}
---. 2013, \apj, 773, 44, \dodoi{10.1088/0004-637X/773/1/44}

\bibitem[{{Wang} {et~al.}(2016){Wang}, {Wu}, {Neri}, {Fan}, {Walter},
  {Carilli}, {Momjian}, {Bertoldi}, {Strauss}, {Li}, {Wang}, {Riechers},
  {Jiang}, {Omont}, {Wagg}, \& {Cox}}]{wang2016}
{Wang}, R., {Wu}, X.-B., {Neri}, R., {et~al.} 2016, \apj, 830, 53,
  \dodoi{10.3847/0004-637X/830/1/53}

\bibitem[{{Wardlow} {et~al.}(2017){Wardlow}, {Cooray}, {Osage}, {Bourne},
  {Clements}, {Dannerbauer}, {Dunne}, {Dye}, {Eales}, {Farrah}, {Furlanetto},
  {Ibar}, {Ivison}, {Maddox}, {Micha{\l}owski}, {Riechers}, {Rigopoulou},
  {Scott}, {Smith}, {Wang}, {van der Werf}, {Valiante}, {Valtchanov}, \&
  {Verma}}]{wardlow2017}
{Wardlow}, J.~L., {Cooray}, A., {Osage}, W., {et~al.} 2017, \apj, 837, 12,
  \dodoi{10.3847/1538-4357/837/1/12}

\bibitem[{{Wei{\ss}} {et~al.}(2003){Wei{\ss}}, {Henkel}, {Downes}, \&
  {Walter}}]{weiss2003}
{Wei{\ss}}, A., {Henkel}, C., {Downes}, D., \& {Walter}, F. 2003, \aap, 409,
  L41, \dodoi{10.1051/0004-6361:20031337}

\bibitem[{{Wei{\ss}} {et~al.}(2005){Wei{\ss}}, {Walter}, \&
  {Scoville}}]{weiss2005}
{Wei{\ss}}, A., {Walter}, F., \& {Scoville}, N.~Z. 2005, \aap, 438, 533,
  \dodoi{10.1051/0004-6361:20052667}

\bibitem[{{Wolfire} {et~al.}(2010){Wolfire}, {Hollenbach}, \&
  {McKee}}]{wolfire2010}
{Wolfire}, M.~G., {Hollenbach}, D., \& {McKee}, C.~F. 2010, \apj, 716, 1191,
  \dodoi{10.1088/0004-637X/716/2/1191}

\bibitem[{{Yang} {et~al.}(2019){Yang}, {Venemans}, {Wang}, {Fan}, {Novak},
  {Decarli}, {Walter}, {Yue}, {Momjian}, {Keeton}, {Wang}, {Zabludoff}, {Wu},
  \& {Bian}}]{yang2019}
{Yang}, J., {Venemans}, B., {Wang}, F., {et~al.} 2019, \apj, 880, 153,
  \dodoi{10.3847/1538-4357/ab2a02}

\bibitem[{{Zanella} {et~al.}(2018){Zanella}, {Daddi}, {Magdis}, {Diaz Santos},
  {Cormier}, {Liu}, {Cibinel}, {Gobat}, {Dickinson}, {Sargent}, {Popping},
  {Madden}, {Bethermin}, {Hughes}, {Valentino}, {Rujopakarn}, {Pannella},
  {Bournaud}, {Walter}, {Wang}, {Elbaz}, \& {Coogan}}]{zanella2018}
{Zanella}, A., {Daddi}, E., {Magdis}, G., {et~al.} 2018, \mnras, 481, 1976,
  \dodoi{10.1093/mnras/sty2394}

\bibitem[{{Zavala} {et~al.}(2018){Zavala}, {Monta{\~n}a}, {Hughes}, {Yun},
  {Ivison}, {Valiante}, {Wilner}, {Spilker}, {Aretxaga}, {Eales},
  {Avila-Reese}, {Ch{\'a}vez}, {Cooray}, {Dannerbauer}, {Dunlop}, {Dunne},
  {G{\'o}mez-Ruiz}, {Micha{\l}owski}, {Narayanan}, {Nayyeri}, {Oteo}, {Rosa
  Gonz{\'a}lez}, {S{\'a}nchez-Arg{\"u}elles}, {Schloerb}, {Serjeant}, {Smith},
  {Terlevich}, {Vega}, {Villalba}, {van der Werf}, {Wilson}, \&
  {Zeballos}}]{zavala2018}
{Zavala}, J.~A., {Monta{\~n}a}, A., {Hughes}, D.~H., {et~al.} 2018, Nature
  Astronomy, 2, 56, \dodoi{10.1038/s41550-017-0297-8}

\bibitem[{{Zavala} {et~al.}(2022){Zavala}, {Casey}, {Spilker}, {Tadaki},
  {Tsujita}, {Champagne}, {Iono}, {Kohno}, {Manning}, \&
  {Monta{\~n}a}}]{zavala2022}
{Zavala}, J.~A., {Casey}, C.~M., {Spilker}, J., {et~al.} 2022, \apj, 933, 242,
  \dodoi{10.3847/1538-4357/ac7560}

\bibitem[{{Zhang} {et~al.}(2016){Zhang}, {Papadopoulos}, {Ivison}, {Galametz},
  {Smith}, \& {Xilouris}}]{z.y.zhang2016}
{Zhang}, Z.-Y., {Papadopoulos}, P.~P., {Ivison}, R.~J., {et~al.} 2016, Royal
  Society Open Science, 3, 160025, \dodoi{10.1098/rsos.160025}

\end{thebibliography}

\end{document}